\documentclass[11pt,x11names,a4paper]{article}

\usepackage[utf8]{inputenc}
\usepackage{lmodern}
\usepackage[T1]{fontenc} 
\usepackage{microtype} 

\usepackage[a4paper, left=25mm, right=25mm, top=30mm, bottom=25mm]{geometry} 

\usepackage{abstract}

\usepackage{xcolor}

\usepackage{cite}
\usepackage{hyperref}
\hypersetup{
	colorlinks=true,
	linkcolor=Blue4,
	citecolor=Red4,
	urlcolor=Green4,
	linktoc=page
}

\usepackage{mathtools}
\usepackage{physics}
 \def\be{\begin{equation}}
 \def\ee{\end{equation}}

\usepackage{tikz}
\usetikzlibrary{decorations.markings,positioning}

\usepackage{amsmath,amssymb,slashed,mathbbol,mathtools}
\numberwithin{equation}{section}

\title{\fontsize{20pt}{24pt}\selectfont\textbf{M5 Instantons in ABJM}\vspace{2mm}}

\author{
\large{\href{mailto:jieming.lin22@imperial.ac.uk}{Jieming Lin}, \href{mailto:jvanmuid@ic.ac.uk}{Jesse van Muiden}, and \large{\href{mailto:zihan.wang18@imperial.ac.uk}{Zihan Wang}}}\\[5mm]
{}{Abdus Salam Centre for Theoretical Physics, Imperial College London}\\
{\normalsize Prince Consort Road, London SW7 2AZ, UK}\\[5mm]
{\small\texttt{jieming.lin22@imperial.ac.uk}, \texttt{jvanmuid@ic.ac.uk}, \texttt{zihan.wang18@imperial.ac.uk} }
}

\date{}

\usepackage{booktabs}
\usepackage{verbatim}

\newcommand{\rmi}{\mathrm{i}}
\newcommand{\rme}{\mathrm{e}}
\newcommand{\rmd}{\mathrm{d}}

\newcommand{\al}[1]{\begin{align}{#1}\end{align}}

\begin{document}
{\hypersetup{urlcolor=black}\maketitle}
\thispagestyle{empty}

\begin{abstract}
\noindent We study the existence of supersymmetric M5-brane instantons in asymptotically AdS$_4 \times S^7/\mathbf{Z}_k$ geometries of M-theory, dual to the ABJM theory. The supersymmetric embeddings of these instantons are determined by the M5-brane $\kappa$-symmetry, which forces them to localise to the equivariant fixed points in the asymptotically AdS$_4$ geometry, and  wrap $(S^3\times S^3)/\mathbf{Z}_k $. At the quantum level we find that the one-loop partition function of these instantons vanishes, suggesting that the bulk partition function is instead to be expanded in terms of supergravity modes and M2-brane instantons, in line with known results for the $S^3$ partition function of the dual ABJM theory. 
\end{abstract}

\newpage

\tableofcontents

\section{Introduction}
Top down holographic dualities provide a direct method to study M-theory, with the three-dimensional ABJM theory and the six-dimensional (2,0) theory as prime examples. Through exact QFT methods, such as supersymmetric localization, the partition function of these theories can be computed as a function of their rank $N$, which are subsequently to be equated to their bulk M-theory counterparts. This has been particularly useful for the $S^3$ and $S^5 \times S^1$ partition functions of the respective theories, which both suggest a bulk expansion into a perturbative supergravity part and M2-brane instantons \cite{Fuji:2011km,Marino:2011eh,Hatsuda:2012dt,Kim:2012ava,Kim:2012qf,Gautason:2023igo,Beccaria:2023ujc,Beccaria:2023sph,Gautason:2025per}, such that
\begin{equation}\label{Eq: holographic equality part funcs}
	\mathcal Z_{\text{M2}}[g,A_3,\Psi] = \log Z_{\text{QFT}}[J]\,.
\end{equation}
The left hand side is to be computed as a functional of the background couplings $(g,A_3,\Psi)$, which have to be correctly mapped to boundary couplings, denoted as $J$. This led recently to the proposal that AdS$_4 \times S^7/\mathbf Z_k$ holography naturally equates bulk dynamics to a grand canonical ensemble of boundary theories at fixed chemical potential $\mu$ (fixing the boundary value of the gauge potential $A_3$), instead of fixed the rank $N$ (the conjugate charge $\star G_4$) \cite{Gautason:2025per,Gautason:2025plx,Gautason:2025bft,vanMuiden:2026nsp}, see also \cite{BenettiGenolini:2026cyc}.\footnote{Similar ensemble structures are expected to exist in type II geometries as well, see for example \cite{vanMuiden:2026lno} for a recent precision test of these ensembles in AdS$_5 \times S^5$, and also \cite{Adhikari:2026xdb}.} Precision holography at fixed $N$ requires a subsequent ensemble changing integral transform
\begin{equation}\label{Eq: Ensemble change}
	\rme^{\mathcal Z_{\text{M5}}[\pi_7]} = \int [\mathcal D a_3] \,\rme^{\mathcal Z_{\text{M2}}[a_3]} \,\rme^{\langle a_3, \pi_7\rangle}\,,
\end{equation}
where $a_3$ is the boundary value of the gauge potential, $\pi_7$ its canonical conjugate which in our setup is simply $\star G_4$, and the measure of this integral is sensitive to the spacetime topology \cite{Bobev:2026gir}. The left had side was denoted with $\mathcal Z_{\text{M5}}$ as it  is a functional of the background six-form potential, not the three-form. 

The superconformal index of the ABJM theory on the other hand was shown to admit an expansion into M5-brane giant gravitons directly in the fixed $N$ ensemble \cite{Arai:2020uwd,Gaiotto:2021xce,Beccaria:2023cuo}, which begs the question if other supersymmetric partition functions of the ABJM theory have non-trivial M5-brane contributions or can even be rewritten as an M5-brane expansion all together. 

Motivated by this discussion we study the existence of semi-classical M5-brane instantons in AdS$_4 \times S^7/\mathbf Z_k$, for a variety of supersymmetric boundary conditions, including empty AdS, supersymmetric thermal AdS, and the supersymmetric Reisner-Nordstrom and Kerr-Newman AdS black holes. We use the PST action to quantise the 5-branes \cite{Pasti:1997gx}. Through a careful $\kappa$-symmetry analysis we show that indeed there are supersymmetric M5-brane instantons, embedded along $(S^3 \times S^3)/\mathbf{Z}_k \subset S^7/\mathbf{Z}_k$, and that they get equivariantly localised according to the background Killing spinors. This result is very much analog to the findings in \cite{Gautason:2025per} regarding the equivariant localisation of M2-brane instantons.

The existence of these M5-brane isntantons however rely on non-trivial worldvolume flux, and as such their origins come from polarized M2-branes \cite{Myers:1999ps,Bena:2000zb}. In particular, in the tower of instantons these five-branes are to be understood as an effective description of a stack M2-brane instantons whose classical action scales as
\begin{equation}
	\rme^{-n S_{\text{M2}}} \sim \rme^{-S_{\text{M5}}} \quad \text{for} \quad n\sim \mu\,,
\end{equation}
where in Planck units the chemical potential $\mu \sim (L/\ell_p)^3$. The world-volume flux shows that the five-branes come from a combination of M2-branes whose type IIA reductions are given by both D2-branes and worldsheet instantons. After classifying the supersymmetric M5-brane instantons we move on to their quantisation. Due to the non-trivial worldvolume flux the 5-brane fluctuations are governed by an effective open membrane metric \cite{Bergshoeff:2000jn,Gibbons:2000ck,VanderSchaar:2001ay,Berman:2001rka,Bergshoeff:2001xx} 
\begin{equation}
	g^{\text{OM}}_{\mu\nu} \propto P[g_{\mu\nu}] - H_{\mu\rho\sigma}H_{\nu}^{\phantom{\nu}\rho\sigma}\,,
\end{equation}
where $P[\cdot]$ denotes the pull-back onto the world-volume of the brane. In particular, this open membrane metric is not the standard round Euclidean metric on $(S^3 \times S^3)/\mathbf{Z}_k$, but instead has a mixed signature $(+++---)$. Even though the 5-brane embedding is classically supersymmetric, this open membrane metric imposes an complications at the quantum level with an infinite number of bosonic and fermionic zero-modes, due to cancellations of non-zero modes between the two spheres, and due to the non-trivial three cycles one is to sum over flux sectors of the chiral three-form \cite{Witten:1996hc,Belov:2006jd}. We deal with the cancelling modes by analytically continuing the effective metric on the M5-brane worldvolume in the complex plane and subsequently find that the one-loop partition function vanishes. Our results suggest that the bulk partition function does not get any contribution from M5-branes, obstructing a possible direct bulk computation of the lefthand side in \eqref{Eq: Ensemble change}, at least in our semi-classical approach.

Finally, as an independent cross-check and confirmation of our results we show how the M5-brane instanton in $\text{AdS}_4 \times S^{7}/\mathbf{Z}_k$ is directly related to defect M5-branes studied in $\text{AdS}_7 \times S^4$ \cite{Lunin:2007ab,Beccaria:2024gkq} through an analytic continuation of the background. 

The remainder of this paper is organised as follows. After introducing the M-theory backgrounds considered in this work in Section~2, we construct the supersymmetric M5-brane instantons and determine their induced M2-brane charges in Section~3. In Section~4, we derive the quadratic worldvolume fluctuations. Given the fluctuations, in Section~5, we evaluate the one-loop partition function and shows that the M5-brane-instanton contribution vanishes for general $k$. Finally, the appendices discuss the analytic continuation relating our solutions to defect M5-branes in AdS$_7\times S^4$ and collect the relevant Killing spinors.

\section{M-theory Backgrounds}\label{Sec: backgrounds}
To study the semi-classical quantisation of five-branes we first fix the asymptotically AdS M-theory background, determined in the semi-classical limit as saddle points of the action
\begin{equation}
	S[g,A_3] = - \frac{2\pi}{(2\pi \ell_p)^9} \int \left( \star (R - \frac12 |G_4|^2) + \frac{\rmi}{6} A_3 \wedge G_4 \wedge G_4 \right)\,.
\end{equation}
Explicitly evaluating this action we do with Dirichlet boundary conditions for the metric and the three-form gauge potential, and add the GHY term. For the three-form we fix the gauge invariant quantity on the AdS$_4$ boundary 
\begin{equation}\label{Eq: definition of mu}
	\exp(\mu) = \exp \left( \frac{2\pi \rmi}{(2\pi \ell_p)^3} \int a_3 \right) = \exp(\mathcal F L^3/\ell_p^3)\,,
\end{equation}
where $a_3$ is the pullback of the gauge potential to the boundary and $\mathcal F$ depends on the particular asymptotic boundary conditions, and will be defined shortly. Explicitly plugging the equations of motion back into the action one finds that its on-shell value can be calculated as a total derivative
\begin{equation}
	S_{\text{on-shell}} = \frac{1}{3}\frac{2\pi}{(2\pi \ell_p)^9} \int \rmd(A_3 \wedge \star G_4) + S_{\text{GHY}}\,.
\end{equation}
The class of solutions we study is found from uplifting solutions of four-dimensional gauged supergravity. More specifically, the backgrounds we study all take the form of a warped product $\mathcal M_4 \times_w S^7/\mathbf{Z}_k$, where $\mathbf{Z}_k$ is freely acting and $\mathcal M_4$ has asymptotically local AdS$_4$ boundary conditions. The internal manifold has locally a $\mathbf{CP}^3$ base that we use to write the consistent truncation ansatz \cite{Gauntlett:2007ma}
\begin{equation}\label{Eq: general 11d background}
	\rmd s_{11}^2 = L^2(\rmd s_{4}^2 + 4 \rmd s_{\textbf{CP}^3}^2 + (\rmd y + \Sigma - A/2)^2 )\,,\qquad A_3 = \rmi L^3(3\omega_3 + \tfrac12\star_4 F \wedge \Sigma)\,,
\end{equation}
where $\rmd \omega_3 = \text{vol}_4$ is the four-dimensional volume form, $F = \rmd A$ is a four-dimensional flux, and $\rmd \Sigma = 4 J$ with $J$ the K\"ahler form on $\mathbf{CP}^3$. This bosonic background solves the eleven-dimensional equations of motion as long as the four-dimensional metric and gauge field solve the equations of motion of $\mathcal N=2$ minimal gauged supergravity, which equal
\begin{equation}
	0 = R_{\mu\nu}^{(4)} + 3g_{\mu\nu}^{(4)} - \frac12 (F_{\mu\rho} F_\nu^{\phantom{\nu}\rho} -\frac14 g_{\mu\nu }^{(4)} F_{\rho\sigma}F^{\rho\sigma} )\,,\quad 0 = \rmd \star_4 F\,.
\end{equation}
Plugging the ansatz back into the eleven-dimensional action it evaluates to 
\begin{equation}\label{Eq: 4d action and cal F}
	S_{\text{on-shell}} = -\frac{2 \text{vol}_7}{\pi^6} \frac{L^9}{\ell_p^9} \mathcal F[F] \,, \quad \text{where}\quad \mathcal F[F] = \frac{1}{(2\pi)^2} \int\limits_{\mathcal M_4} \star_4 \left( 3 + \frac{1}{8} F_{\mu\nu}F^{\mu\nu} \right)\,.
\end{equation}
Here, $\text{vol}_7$ is the volume of the unit-radius $S^7/\mathbf{Z}_k$. As explained in \cite{Bobev:2026gir} it is this parameter $\mathcal F$ which feeds into the background value for $\mu$ in \eqref{Eq: definition of mu}. To study the M5-brane instantons in this background it will be useful to introduce explicit coordinates on the internal seven-sphere, specify how the orbifold acts, and present the background Killing spinors. First, the freely acting $\mathbf Z_k$ orbifold acts on the internal seven-sphere through
\begin{equation}
	\mathbf Z_k:\quad (z_1,z_2,z_3,z_4) \rightarrow (\rme^{2\pi \rmi/k} z_1,\rme^{2\pi \rmi/k} z_2,\rme^{-2\pi \rmi/k} z_3,\rme^{-2\pi \rmi/k} z_4)\,,
\end{equation}
where $z_i$ are the $\mathbf{C}^4$ embedding coordinates of the seven-sphere. We will parametrise the seven-sphere as a fibration of two three-spheres over a line such that
\begin{equation}
	(z_1,z_2) = \sin \theta (u_1^\psi,u_2^\psi)\,,\qquad (z_3,z_4) = \cos \theta (u_1^\eta,u_2^\eta)\,,
\end{equation}
where the $u_i^x$ coordinates span two three-spheres
\begin{equation}
	u_1^x = \frac{1}{\sqrt{2}}\rme^{\rmi (x_1 + x_3)} (\cos x_2  - \sin x_2)\,,\qquad u_2^x = \frac{1}{\sqrt{2}}\rme^{\rmi (x_1 - x_3)} (\cos x_2  + \sin x_2)\,. 
\end{equation}
And subsequently the orbifold acts non-trivially on $\mathbf{CP}^3$, identifying
\begin{equation}
	(\psi_1,\eta_1)\,\sim \, (\psi_1 + \frac{2\pi}{k}, \eta_1 - \frac{2\pi}{k})\,,
\end{equation}
We choose the frames on the internal space accordingly
\begin{equation}\label{Eq: frames on seven sphere}
\begin{aligned}
	\rme^{5,\ldots,11} = 2L(\rmd \theta, \,\sin\theta\, \sigma^\psi_{1,2,3}, \,\cos\theta \, \sigma^\eta_{1,2,3}) \,,
\end{aligned}
\end{equation}
where
\begin{equation}\label{Eq: fibration of graviphoton over three spheres}
	\sigma_1^x = -\frac{\rmi}{2}\sum_{i=1}^2 (\bar u_i^x \rmd u_i^x - u_i^x \rmd \bar u_i^x) - A/4 \,,\qquad \sigma_2^x - \rmi \sigma_3^x = u_1^x \rmd u_2^x - u_2^x \rmd u_1^x\,.
\end{equation}
The Hopf fibre in \eqref{Eq: general 11d background} is parametrized with $y = \psi_1 + \eta_1$ and the K\"ahler form equals 
\begin{equation}
	\Sigma = \cos 2\theta \,(\dd \eta_1-\dd \psi_1) - 2\cos^2\theta\, \sin 2\eta_2\, \dd \eta_3 - 2\sin^2\theta \,\sin 2\psi _2 \, \dd \psi_3 \,, \qquad 4J = \rmd \Sigma\,.
\end{equation}
The bosonic consistent truncation to minimal gauged supergravity, as described above, extends to the spinors as well, upon taking the ansatz for the background Killing spinors to be\cite{Gautason:2025per}\footnote{The eleven-dimensional Gamma-matrices are decomposed in a 4+7 split as 
\begin{equation}
	\Gamma_M = (\rmi \gamma_{(4)} \gamma_\mu \otimes \mathbb 1,-\gamma_{(4)} \otimes  \Sigma_m)\,, \quad \gamma_{(4)}=\gamma_{1}\gamma_{2}\gamma_{3}\gamma_{4}.
\end{equation}}
\begin{equation}\label{Eq: consistent truncation of 11d Killing spinor}
	\epsilon_{11} = \epsilon_4 \otimes \chi^+ + \epsilon_4^c \otimes \chi^-\,,
\end{equation}
where $\chi^{\pm}$ are Killing spinors on $S^7/\mathbf{Z}_k$ 
\begin{equation}\label{Eq:Chi_pm}
		\chi^{\pm} =  \rme^{\rmi \frac \theta 2\Sigma_{5}}\frac{1 \pm \rmi \Sigma_{7\,8}}{2}\,  \frac{1 - \rmi \Sigma_{5\,6\,9}}{2} \,\frac{1 + \rmi \Sigma_{9\,10\,11}}{2} \chi_0\,.
\end{equation}
This split of the Killing spinors is particularly useful as they naturally project onto the eigenspaces of the K\"ahler form 
\begin{equation}
	\frac{1}{3} (\frac{\rmi}{2} J_{mn} \Sigma^{mn}) \chi^{\pm} = \pm \chi^{\pm}\,.
\end{equation}
Indeed, with this ansatz for the Killing spinors the variation of the eleven-dimensional gravitino
\begin{equation}\label{Eq: 11d BPS equation}
	\delta \Psi_M = \left( \nabla_M + \frac{1}{24} \Gamma_M \slashed{G}_4 - \frac18 \slashed{G}_4 \Gamma_M \right) \epsilon_{11} = 0\,
\end{equation}
factorises. The variations along the internal directions vanish and one is left with solving the equations along the four-dimensional space which reduce to the variations of the four-dimensional gravitino:
\begin{equation}
	\delta_{\epsilon} \Psi_\mu = \delta_{\epsilon_4} \psi_\mu \otimes \chi^+ + \delta_{\epsilon_4^c} \bar \psi_\mu \otimes \chi^-\,,
\end{equation}
where 
\begin{equation}
	\delta_{\epsilon_4} \psi_\mu = \nabla_\mu \epsilon_4 - \frac{\rmi}{2} A_\mu \epsilon_4 + \frac12 \gamma_\mu \epsilon_4 + \frac{\rmi}{4} \slashed{F} \gamma_\mu \epsilon_4\,.
\end{equation}
Before highlighting a set of explicit four-dimensional saddles of interest to us, we first quickly recall the recent results simplifying the evaluation of on-shell actions in supersymmetric backgrounds using the Berline-Vergne-Atiyah-Bott fixed point formula \cite{BenettiGenolini:2023kxp}. It was shown that the action integral equivariantly localizes to contributions from fixed points in the geometry which are in one-to-one with either isolated nuts or two-dimensional bolts where the Killing spinor turns chiral
\begin{equation}\label{Eq: 4d chirality constraint}
	\gamma_{(4)} \epsilon_4 = \pm \epsilon_4\,.
\end{equation}
Beyond the two-derivative approximation it was shown that this equivariant localization procedure is valid also for the eight-derivative topological term in eleven-dimensional supergravity \cite{BenettiGenolini:2026cdw}, and furthermore that the non-perturbative sectors coming from membrane instantons localize to these chiral points as well \cite{Gautason:2025per}. As we present below, the same equivariant localization is valid for the non-perturbative M5-brane instantons in supersymmetric asymptotically AdS$_4$ geometries.

The M5-brane instanton analysis will be universal for supersymmetric asymptotically locally AdS solutions of four-dimensional minimal $\mathcal N=2$ gauged supergravity uplifted on the seven sphere. It will turn out that the fluctuations on the brane are fixed by the four-dimensional graviphoton evaluated at the BPS locus, denoted $\bar x$, of the brane. At this point we split the graviphoton flux into two planes
\begin{equation}\label{Eq: fixed point flux weights}
	F(\bar x)=\omega_{1}\,\rme^1\wedge\rme^2+\omega_{2}\,\rme^3\wedge\rme^4\,,\qquad f^\pm\equiv\frac12\left(\omega_{1} \pm \omega_{2}\right)\,.
\end{equation}
Thus $f^+$ and $f^-$ are respectively the self-dual and anti-self-dual weights in this orientation. For concreteness we will now list five particularly interesting solutions to which our results apply and provide explicit evaluations of the graviphoton. 

\paragraph{Empty AdS.} The simplest geometry is empty AdS with a spherical boundary: 
\begin{equation}
	\rmd s_4^2 = \rmd \rho^2 + \sinh^2 \rho (\rmd \psi^2 + \cos^2 \psi \rmd \phi_1 ^2 + \sin^2 \psi \rmd \phi_2^2)\,,\quad A = 0\,.
\end{equation}
Holographically this geometry corresponds to the sphere partition function of the ABJM theory. The isolated fixed point is the nut at $\rho=0$. The Killing spinor is chiral there, with either chirality available, and the vanishing graviphoton gives 
\begin{equation}
	\rho=0:\qquad \gamma_{(4)}\epsilon_4=\pm\epsilon_4\,,\qquad (f^+,f^-)=(0,0)\,.
\end{equation}

\paragraph{Anti-self-dual instanton.} The metric in this case is the same as in empty AdS, but the background includes a non-trivial anti-self-dual instanton \cite{Martelli:2011fu}. Explicitly, with the following set of coordinates the background takes the form
\begin{equation}
\begin{aligned}
	\rmd s_4^2 =&\, \rmd \rho^2 + \sinh^2 \rho (\rmd \psi^2 + \cos^2 \psi \rmd \phi_1 ^2 + \sin^2 \psi \rmd \phi_2^2)\,,\\
	A =&\, \frac{(b^2 + \cosh \rho)\rmd \phi_2 - (1 + b^2 \cosh \rho)\rmd \phi_1}{\sqrt{S_+ S_-}} + \rmd \phi_1 - \rmd \phi_2\,,	
\end{aligned}
\end{equation}
where
\begin{equation}
	S_{\pm} = {(b^2 + \cosh \rho)\cos\psi \pm \rmi(1 + b^2 \cosh \rho)\sin\psi}\,,
\end{equation}
and the gauge field is chosen to be globally well defined. We have chosen coordinates such that the metric is that of empty AdS, however, the boundary geometry is that of a squashed three sphere $S^3_b$, with $b$ the conventional $\text{U}(1) \times \text{U}(1)$ squashing parameter \cite{Hama:2011ea}. The isolated fixed point is again the nut at $\rho=0$, but only one chirality survives. With the positive orientation used in \eqref{Eq: fixed point flux weights}, the anti-self-dual graviphoton has 
\begin{equation}
	\rho=0:\qquad \gamma_{(4)}\epsilon_4=\epsilon_4\,,\qquad
	(f^+,f^-)=\left(0,\frac{1-b^2}{1+b^2}\right)\,.
\end{equation}

\paragraph{Thermal AdS.} This is simply global AdS, where the compactified time is twisted with an internal $\text{U}(1)$ isometry to ensure that the Killing spinors are globally well defined along the time circle, and supersymmetry is preserved. The four-dimensional metric and flux equal
\begin{equation}
	\rmd s_4^2 = \rmd \rho^2 + \cosh^2 \rho \, \rmd \tau^2 + \sinh^2 \rho \,\rmd s_{S^2}^2\,,\quad A=0\,.
\end{equation}
The twist is a global identification in the uplift, rather than a non-trivial local four dimensional field strength, and so we take $A=0$. We will show below that there are in fact no brane instantons, fully embedded in the internal geometry, in this background. Instead the non-perturbative corrections arise entirely from five-brane giant gravitons \cite{Arai:2020uwd}. The time circle remains finite in the interior, so the Killing spinor has no chiral fixed locus and there are no non-trivial BPS instanton saddles \cite{Gautason:2025per}. Holographically, this geometry counts states contributing to the superconformal index whose dimensions scale as $\mathcal O(1)$ or $\mathcal O(N)$, with the latter coming from M5-giants. 

\paragraph{Dyonic black hole.} The dyonic black hole solutions in asymptotically AdS$_4$ geometries can have Riemann surface horizons and locally the frame fields can be chosen to be \cite{Romans:1991nq,Caldarelli:1998hg} 
\begin{align}\label{dyonmetric}
	&e^1 = V^{-1/2}\rmd \rho \,,\quad e^2 = V^{1/2}\rmd  \tau\,,\quad e^3 = \rho \,\rmd x \,,\quad e^4 = \rho \,\sinh x\,\rmd y \,.
	\\
	&V=V_+ V_- \,,\quad V_{\pm} = \rho - \frac{1\pm 2Q}{2\rho}\,,
\end{align}
where the Riemann surface has genus $g>1$ and is equipped with a constant negative-curvature metric. The gauge field equals
\begin{equation}
	A = -\frac{2Q}{\rho} \rmd \tau + \omega_{\Sigma_g}\,, \quad \omega_{\Sigma_g}=-\cosh x \rmd y
\end{equation}
The supersymmetric Killing vector vanishes on the bolt $V_+=0$, which is the black-hole horizon at $\rho_+^2=(1+2Q)/2$, where
\begin{equation}
	F\big|_{\rho_+}=\frac{4Q}{1+2Q}\,\rme^1\wedge\rme^2-\frac{2}{1+2Q}\,\rme^3\wedge\rme^4\,.
\end{equation}
The Killing spinor therefore has definite chirality on this bolt, and
\begin{equation}
	V_+=0:\qquad \gamma_{(4)}\epsilon_4=\epsilon_4\,,\qquad
	(f^+,f^-)=\left(\frac{2Q-1}{2Q+1},1\right)\,.
\end{equation}
\paragraph{Rotating black hole.}
The electrically charged Kerr-Newman-AdS$_4$ black hole gives the rotating solution of minimal gauged supergravity \cite{Carter:1968ks,Kostelecky:1995ei,Caldarelli:1998hg,Cassani:2019mms}.  In Euclidean signature, and in units where the AdS radius is one, the Kerr-Newman-AdS$_4$ black hole metric can be written as
\begin{equation}
\begin{aligned}
	\rmd s_4^2 =&\, \mathcal S\left(\frac{\rmd r^2}{\Delta_r}+\frac{\rmd\theta^2}{\Delta_\theta}\right)+ \frac{\Delta_r}{\mathcal S}\left(\rmd \tau + \frac{a\sin^2\theta}{1 + a^2} \rmd\phi\right)^2 +\frac{\Delta_\theta\sin^2\theta}{\mathcal S}\left(a\rmd \tau-\frac{r^2-a^2}{1 + a^2}\rmd\phi\right)^2\,,\\
	A =&\, \frac{2qr}{\mathcal S}\left(\rmd \tau + \frac{a\sin^2\theta}{1 + a^2}\rmd\phi\right)\,,
\end{aligned}
\end{equation}
where
\begin{equation}
	\mathcal S = r^2 - a^2\cos^2\theta\,,\qquad
	\Delta_r=(r^2 - a^2)(1 + r^2) - 2mr - q^2\,,\qquad
	\Delta_\theta=1 + a^2\cos^2\theta\,.
\end{equation}
The supersymmetric branch obeys
\begin{equation}\label{Eq: KN susy}
	q = -\frac{\rmi m}{1 + \rmi a}\,,
\end{equation}
and the extremal branch imposes additionally
\begin{equation}\label{Eq: KN extremal}
	m^2 = \rmi a(1 + \rmi a)^4\,,\qquad q^2 = -\rmi a(1 + \rmi a)^2\,,\qquad r_+^2 = \rmi a\,.
\end{equation}
Equivalently, writing $a = -\rmi \alpha$ the regular supersymmetric extremal branch of solutions is 
\begin{equation}\label{Eq: KN BPS}
	a = -\rmi \alpha\,,\qquad m = \sqrt{\alpha}(1 + \alpha)^2 \,,\qquad q = -\rmi \sqrt{\alpha}(1 + \alpha)\,,\qquad r_+ = \sqrt{\alpha}\,,
\end{equation}
where $0<\alpha<1$.
The horizon sits at the double root $\Delta_r(r_+)=\Delta_r'(r_+)=0$. In the Euclidean Carter--Pleba\'nski description of this family, the supersymmetric Killing vector has two isolated fixed points. For the Kerr--Newman specialization these are precisely the north and south poles of the horizon,
\begin{equation}
	(r,\theta)=(r_+,0)\,,\qquad (r,\theta)=(r_+,\pi)\,.
\end{equation}
And the Killing spinor has definite chirality at each of these fixed points. To state the flux weights, define
\begin{equation}
\begin{aligned}
	\rme^1&=\sqrt{\frac{\mathcal S}{\Delta_r}}\,\rmd r\,,\qquad
	\rme^2=\sqrt{\frac{\Delta_r}{\mathcal S}}\left(\rmd\tau+\frac{a\sin^2\theta}{1+a^2}\rmd\phi\right)\,,\\
	\rme^3&=\sqrt{\frac{\mathcal S}{\Delta_\theta}}\,\rmd\theta\,,\qquad
	\rme^4=-\sqrt{\frac{\Delta_\theta}{\mathcal S}}\sin\theta\left(a\rmd\tau-\frac{r^2-a^2}{1+a^2}\rmd\phi\right)\,.
\end{aligned}
\end{equation}
for which the two graviphoton blocks are
\begin{equation}
	F=-\frac{2q(r^2+a^2\cos^2\theta)}{\mathcal S^2}\,\rme^1\wedge\rme^2
	+\frac{4qra\cos\theta}{\mathcal S^2}\,\rme^3\wedge\rme^4\,.
\end{equation}
Consequently, at the north and south fixed points, respectively,
\begin{equation}
\begin{aligned}
	(f^+,f^-)_{\mathrm N}
	&=\frac{\rmi}{\sqrt\alpha}\left(\frac{1+\rmi\sqrt\alpha}{1-\rmi\sqrt\alpha},
	\frac{1-\rmi\sqrt\alpha}{1+\rmi\sqrt\alpha}\right)\,,\\
	(f^+,f^-)_{\mathrm S}
	&=\frac{\rmi}{\sqrt\alpha}\left(\frac{1-\rmi\sqrt\alpha}{1+\rmi\sqrt\alpha},
	\frac{1+\rmi\sqrt\alpha}{1-\rmi\sqrt\alpha}\right)\,.
\end{aligned}
\end{equation}

\section{M5-brane instantons}
Having set up the supergravity saddles of interest in M-theory, we will now move on to study the BPS M5-brane instantons arising on top of them. We analyse these instantons in the PST formalism and have the following four contributions to the action \cite{Pasti:1997gx,Bandos:1997ui,Aganagic:1997zq,Howe:1997fb,Bandos:1997gm,Kallosh:2005yu}
\begin{equation}\label{Eq: M5 action}
\begin{aligned}
	S_{\text{M5}}^{}
	& = T_{\text{M5}} (S_{\text{DBI}} + S_{\text{PST}} + S_{\text{WZ}} + S_{\text{GS}})\,,\qquad
	T_{\text{M5}} = \frac{2\pi}{(2\pi \ell_p)^6}\,.
\end{aligned}
\end{equation}
where
\begin{equation}\label{Eq: PST action}
\begin{aligned}
	S_{\text{DBI}} =&\,\int \rmd^6 \xi\,\sqrt{|\hat g _6+ \tilde H_2|}\,,\\
	S_{\text{PST}} =&\, -\frac{i}{4\times 3!} \int \rmd^6\xi\,\varepsilon^{ijklmn} H_{lmn} {H}_{jkp} V_i V^p	\,,\\
	S_{\text{WZ}} =&\, -i q_{\text{M5}} \int \left(\hat A_6 - \frac i2 \mathcal F_3 \wedge \hat A_3\right)\,,
\end{aligned}
\end{equation}
$q_{\text{M5}} = \pm 1$ is the Wess-Zumino orientation, and the Green-Schwarz contribution will be spelled out below.\footnote{We take $\varepsilon^{123456} = +1$.} The worldvolume coordinates are denoted with $\xi_i$ and the pull-backs of the target-space fields, denoted with hats, encode the five dynamical scalar fields on the brane $X^M(\xi)$. Additionally there is a dynamical two-form potential with field strength $\mathcal F_3 = \rmd B_2$, and an auxiliary scalar $a$ out of which we construct the auxiliary vector field $V_i = \partial_i a/\sqrt{(\partial a)^2}$. This auxiliary scalar is added in the PST formalism to ensure covariance of the action with respect to diffeomorphisms on the worldvolume of the five-brane. The usual procedure is to gauge fix it to a particular coordinate, which is what we will do as well. Out of the two-form and the auxiliary scalar we construct the following flux quantities:
\begin{equation}\label{32}
	\tilde H_2 = i_V (\star {H_3})\,,\quad {H}_3 = \mathcal F_3 - \hat{A}_3\,,\quad \mathcal F_3 = \rmd B_2\,,
\end{equation}
Importantly, from varying the action with respect to $B_2$ one finds a non-linear duality constraint on the three-form $H_3$
\begin{equation}\label{Eq: duality constraint}
	 i_{V}H_3=-\rmi\,\mathcal{V}\,,\qquad\mathcal{V}^{mn}=-\frac{\sqrt{|\hat g_6|}}{\sqrt{|\hat g_6+\tilde H_2|}}\left(\left(1+\frac{1}{2}\tilde H_{ij}\tilde H^{ij}\right)\tilde{H}^{mn}+\tilde{H}^{mi}\tilde{H}_{ij}\tilde{H}^{jn}\right)\,.
\end{equation}
To quadratic order in the fermions, the flux-dressed Green-Schwarz term equals \cite{Howe:1997fb,Bandos:1997gm,Kallosh:2005yu}
\begin{equation}\label{Eq: Fermionic action}
\begin{aligned}
	S_{\text{GS}} =& \frac14\int \rmd^6\xi\sqrt{|\hat g_6|}\;
	\bar\Theta\,m^{ij}\Gamma_jD_i (1- \Gamma_*)\Theta\,,
\end{aligned}
\end{equation}
where
\begin{equation}
	m_i{}^j = \delta_i{}^j-2t_{ikl}t^{jkl} \,,\quad  D_i = \partial_iX^M (\nabla_M+\frac1{24}\Gamma_M\slashed{\hat{G}}_4
	-\frac18\slashed{\hat{G}}_4\Gamma_M)\,,
\end{equation}
the $\Gamma$-matrix projector equals
\begin{equation}
\begin{aligned}
	1 - \Gamma_* =&\,  (1 - \Gamma^{(6)})(1 - \frac{1}{3!} t^{ijk} \Gamma_{ijk} )\,,\qquad \Gamma^{(6)}=\frac{\rmi}{6!} \varepsilon^{ijklmn}\Gamma_{ijklmn}
\end{aligned}
\end{equation}
and $t^{ijk}$ is self-dual and non-linearly related to the worldvolume three-form flux
\begin{equation}
	 4 t^{ijk}V_k =  \tilde H^{ij} - \frac{\tilde H^{im}\tilde H_{mn}\tilde H^{nj} - \frac14  \tilde H_{mn} \tilde H^{mn} \tilde H^{ij}}{1 + \frac14 \tilde H_{mn}\tilde H^{mn} + \sqrt{|\hat{g}_6+\tilde{H}_2|}/ \sqrt{\det\hat g_6} }\,,
\end{equation}
To fix the fermionic gauge for the worldvolume fermion we take
\begin{equation}\label{eq:kappa-gauge}
	(1+\Gamma_*) \Theta = (1 + \frac{1}{3!} t^{ijk} \Gamma_{ijk} )(1 + \Gamma^{(6)})\Theta = 0,
\end{equation} 
which for regular background flux configurations is equivalent to the constraint 
\begin{equation}\label{eq:kappa-gauge-result}
	(1 + \Gamma^{(6)}) \Theta = 0\,.
\end{equation}
The matrix $\Gamma_*$ in the quadratic Dirac operator should not be identified with the standard covariant $\kappa$-symmetry matrix of the PST brane. The former is the flux-dependent matrix entering the gauge-fixed fermion kinetic term, whereas the latter is the local fermionic symmetry projector used to test or gauge-fix target-space supersymmetry \cite{Bandos:1997gm,Kallosh:2005yu}. The latter takes the form
\begin{equation}\label{Eq:kappa_projection}
\begin{aligned}
	\Gamma_\kappa \epsilon &= q_{\text{M5}} \epsilon\,,\\
	\Gamma_\kappa & = \frac{\rmi \slashed V}{\sqrt{|\hat g_6+\tilde H_2|}}
	\left[\sqrt{\det\hat g_6} \, \tilde{\slashed H}_2 - \varepsilon^{i j k l m n}V_i \left(\frac{1}{5!}\Gamma_{jklmn} + \frac{1}{2^3}\Gamma_j\tilde H_{kl}\tilde H_{mn}\right)\right]\,.
\end{aligned}
\end{equation}

\subsection{Supersymmetric embeddings}\label{Sec: embeddings}

An M5 solution contains the embedding of world-volume coordinates $X^M(\xi)$, the world-volume 3-form flux $H$ and a choice of gauge on the auxiliary PST scalar $a$. We consider static M5-branes which are wrapped entirely in the internal geometry, and sit at the fixed BPS locus in the asymptotically AdS$_4$ geometry. We work in static gauge and take the embedding to be of the form
\begin{equation}
	\xi_i = (\psi_1,\psi_2,\psi_3,\eta_1,\eta_2,\eta_3)\,,
\end{equation}
and have the brane sit at a fixed value of $\theta$. In what follows we simply denote the worldvolume coordinates with the same labels as the target space coordinates. The worldvolume geometry of this static 5-brane is
\begin{equation}
	(S^3_\psi \times S^3_\eta)/\mathbf{Z}_k\,,
\end{equation}
where again the orbifold simply identifies
\begin{equation}
	(\psi_1,\eta_1) \sim \left( \psi_1 + \frac{2\pi}{k}, \eta_1 - \frac{2\pi}{k} \right)
\end{equation}
and we choose to cover the entire manifold with the following coordinate ranges
\begin{equation}
	0\leq \psi_1 < \frac{2\pi}{k}\,,\qquad 0\leq \eta_1 \leq 2\pi\,,\qquad -\frac{\pi}{4} \leq \psi_2,\eta_2\leq \frac{\pi}{4}\,,\quad 0\leq \psi_3, \eta_3,< \pi\,.
\end{equation}
The induced metric subsequently takes the simple form
\begin{equation}
	\rmd \hat s_{6}^2 = 4L^2\sum_{i=1}^3 \left[\sin^2 \theta (\sigma^\psi_i)^2 + \cos^2 \theta (\sigma^\eta_i)^2\right]\,,
	\label{eq:induced-metric}
\end{equation}
where $\theta$ is still to be fixed. Next, to solve the worldvolume equations of motion we gauge fix the PST scalar to equal\footnote{A symmetrically equivalent choice would be $a = \psi_2$, and we will comment on the results in this case throughout.}
\begin{equation}\label{Eq: PST gauge}
	a = \eta_2
\end{equation}
and make the following symmetry-inspired ansatz for the three-form flux
\begin{equation}
	H_3 = 8 L^3 \alpha(X) \sin^3 \theta\, \sigma_1^\psi \wedge \sigma_2^\psi \wedge \sigma_3^\psi + 8 L^3 \beta(X) \cos^3 \theta\, \sigma_1^\eta \wedge \sigma_2^\eta \wedge \sigma_3^\eta\,,
\end{equation}
where the functions $\alpha(X)$ and $\beta(X)$ depend on the worldvolume scalars, and are to be solved for. The PST gauge makes the contribution of the three-form flux along $S_\eta^3$ vanish in the five-brane action. Nevertheless, the non-linear self-duality constraint \eqref{Eq: duality constraint} relates the two ansatz functions. With \eqref{Eq: PST gauge} the auxiliary scalar and the two-form $\tilde H_2$ simplify to the following non-trivial components
\begin{equation}
	 V_5=2L\cos\theta \,,\qquad \tilde H_{46}=-4\alpha(X) L^2\cos 2\eta_2 \cos^2 \theta\,,
\end{equation}
and the duality constraint \eqref{Eq: duality constraint} on the flux collapses to
\begin{equation}\label{Eq: self duality constraint}
		\beta(X) = \frac{\rmi\,\alpha(X)}{\sqrt{1+\alpha(X)^2}}\,.
\end{equation}
The Euclidean five-brane action simplifies in this configuration to 
\begin{equation}\label{Eq: reduced action}
\begin{aligned}
S_{\mathrm{M5}}= 64 T_{\text{M5}} L^6 \sin^3 \theta \cos^3 \theta\int \text{vol}_{S^3_\psi \times S^3_\eta}\, \Big[&\sqrt{-(1+\alpha(X)^2)}+\frac12\alpha\beta - q_{\text{M5}} (\tan^3 \theta - \frac32 \frac{\tan\theta}{\cos^2 \theta})\Big]\,.
\end{aligned}
\end{equation}
We have chosen a gauge for $A_6$ that makes the Wess-Zumino term vanish as $\theta \to 0$, where one of the induced three-spheres shrinks.  

Instead of explicitly solving for the variation of the $\theta$-scalar and fixing the form of $\alpha(X)$ and $\beta(X)$ we directly move to solving the $\kappa$-symmetry constraints which do the same job. On our embedding the $\kappa$-projector reduces to
\begin{equation}
		\Gamma_\kappa=\frac{\rmi}{\sqrt{1+\alpha(X)^2}}\left(\alpha(X)  - \Gamma_{6\,7\,8}\right)\Gamma_{9\,10\,11}\,.
\end{equation}
And its solutions when acting on the target space Killing spinor \eqref{Eq: consistent truncation of 11d Killing spinor} are of the form
\begin{equation}\label{Eq: kappa constraint split}
	\Gamma_\kappa \epsilon_{11} = q_{\text{M5}} \epsilon_{11} \quad \Rightarrow \quad	\gamma_{(4)} \epsilon_4 = q_4 \epsilon_{4}\,,\quad \text{and} \quad \Sigma_\kappa \chi^\pm = q_{\text{M5}} \chi^\pm\,,
\end{equation}
where both $q_4$ and $q_{\text{M5}}$ are $\pm 1$.  On the internal geometry the $\kappa$-constraint comes from the seven-dimensional projector
\begin{equation}\label{7dkappa}
		\Sigma_\kappa = \frac{\rmi}{\sqrt{1+\alpha(X)^2}} ( -q_4 \alpha(X) \Sigma_{9\,10\,11} + \rmi \Sigma_5)\, ,
\end{equation}
where we have dualised the gamma matrices in seven dimensions to simplify its form. This $\kappa$-constraint, acting on the Killing spinors (\ref{Eq:Chi_pm}), gives the equality
\begin{equation}
( \rmi q_4 \alpha(X) e^{-\rmi \theta \Sigma_5} + \rmi \Sigma_5)\chi^\pm
=-\rmi \sqrt{1+\alpha(X)^2}\,q_{\text{M5}} \chi^\pm\,,
\end{equation}
where we have used the projector identity
\begin{equation}
	\rmi\Sigma_{9\,10\,11}\chi^\pm = \rme^{-\rmi\theta\Sigma_5}\chi^\pm\,.
\end{equation}
The $\kappa$-constraint is subsequently solved by
\begin{equation}
	q_{\mathrm{M5}}=+1\,,\qquad
	\alpha(X) = - \frac{\rmi\,q_4}{\sin\theta}\,,\qquad
	\beta(X) = -\frac{\rmi\,q_4}{\cos\theta}\,,
	\label{eq:sol-alpha-beta}
\end{equation}
where the last equality arose directly from the self-duality constraint \eqref{Eq: self duality constraint}. For the solution above we have that $1+\alpha(X)^2 = -\cot^2\bar\theta$ is negative, and we need to fix the branch in the square root, which is
\begin{equation}\label{Eq: branch}
	\sqrt{1+\alpha(X)^2} = \rmi \cot \bar\theta\,, \qquad 0<\bar\theta<\pi/2\,.
\end{equation}
The four-dimensional part of the $\kappa$-constraint in \eqref{Eq: kappa constraint split} fixes the 5-brane to sit at the equivariant fixed points discussed around \eqref{Eq: 4d chirality constraint}, analogously to what was found in \cite{Gautason:2025per} for supersymmetric M2-brane instantons. Subsequently, there are supersymmetric five-brane instantons for all locally asymptotically AdS$_4$ geometries discussed in Section \ref{Sec: backgrounds}, except for thermal AdS. Indeed, in the thermal AdS case the supersymmetric five-branes that can contribute non-perturbatively to the partition function are giant gravitons, not instantons.
\subsection{M2-brane charge}\label{Sec: M2 charge}
The supersymmetric five-branes found in the previous section have non-trivial background worldvolume flux, signalling that their origin comes from polarized M2-branes through the dielectric Myers effect \cite{Myers:1999ps}. The M2-brane charge on the worldvolume, measured in units of the M2-brane tension, equals
\begin{equation}
	n_{\text{M2}} = \frac{2\pi \rmi}{(2\pi\ell_p)^3} \int_{\mathcal M_3} H_3\,.
\end{equation}
In the embedding coordinates of the brane the two obvious three-cycle representatives along which we compute this charge are $S^3_\psi$ and $S^3_\eta$. In the fundamental domain chosen above, the $\psi_1$ period is reduced by $1/k$, while the $\eta_1$ period is unchanged, and thus we have
\begin{equation}\label{Eq: two M2 charges}
	n_{\text{M2}}^\psi = \frac{4 q_4}{k}\frac{L^3}{\ell_p^3} \sin^2 \theta\,,\qquad n_{\text{M2}}^\eta =  4 q_4 \frac{L^3}{\ell_p^3} \cos^2 \theta\,.
\end{equation}
Their combinations
\begin{equation}\label{Eq: charge combinations}
	k\,n^\psi_{\text{M2}} - n^\eta_{\text{M2}} = -\,4q_4\frac{L^3}{\ell_p^3}\cos2\theta \,,\qquad
	k\,n^\psi_{\text{M2}} + n^\eta_{\text{M2}} = 4q_4\frac{L^3}{\ell_p^3}\,,
\end{equation}
are the chiral and anti-chiral parts of $\bar H_3$ used below.
Due to the worldvolume three-form being self-dual only one half of the $H^3(\mathcal M_6,\mathbf Z)$ charge lattice is independent. The PST gauge $a = \eta_2$ chosen above is what selects which half: with the PST vector aligned along $S^3_\eta$, the flux through $S^3_\psi$ is the M2-brane charge that is to be fixed as a background source, labelling the instanton and determining the on-shell action
\begin{equation}
	S_{\text{M5}} = \frac{4L^6}{\pi k \ell_p^6}\sin^2 \theta
	= \frac{q_4 n_{\text{M2}}^\psi}{\pi \mathcal F} \mu \,.
	\label{Eq: M5 on-shell}
\end{equation}
Had we instead chosen the PST gauge $a = \psi_2$, the roles of the two spheres would be exchanged. These supersymmetric M5-branes with finite action are expected to contribute to the M-theory partition function. They are subleading to the M2-brane instantons such that schematically we expect the M-theory partition function to be expanded into saddle points as follows
\begin{equation}\label{Eq:M-partition}
	\mathcal Z_{\text{M}}(\mu) = -S_{\text{SUGRA}}(\mu) + \ldots + \sum_{\text{M2s}} \rme^{-S_{\text{M2}}(\mu)} (Z_{\text{1-loop}}^{\text{M2}} + \ldots) +  \sum_{\text{M5s}} \rme^{-S_{\text{M5}}^{}(\mu)} (Z_{\text{1-loop}}^{\text{M5}} + \ldots)\,,
\end{equation}
where remember that $\mu = \mathcal F\, L^3/\ell_p^3$ and $\mathcal F$ is defined in \eqref{Eq: 4d action and cal F} and is background dependent. The M2-brane instanton contributions to the partition function were the focus of the recent works \cite{Beccaria:2023ujc,Gautason:2025per,Kurlyand:2026yke}, where a match with the boundary field theory partition function was found when available. The focus of the subsequent sections will be to determine the one-loop determinant of the five-brane instantons. Before doing so let us comment on the type IIA string interpretation of these five-branes. The IIA picture is understood most easily by first going to the coordinate system
\begin{equation}\label{Eq: IIA coordinates}
	y = \psi_1 + \eta_1\,,\quad \varphi = \psi_1 - \eta_1\,,
\end{equation}
and reduce over the $\varphi$ direction. The $\mathbf Z_k$-invariant three-cycle
\begin{equation}\label{Eq: CNS}
	\mathcal C_{\text{NS}} :\quad \eta_1 = -\psi_1\,,\quad \eta_2 = -\psi_2 \,,\quad \eta_3 = \psi_3 + \pi/2\,,
\end{equation}
descends to a lens space $S^3/\mathbf Z_k$ with its period reduced by $1/k$, and it is along this cycle that the single independent flux is quantised on the quotient. Reduced to ten dimensions the polarized M5-brane is a bound state carrying both F1-string and D2-brane charge. Because only one worldvolume flux is independent, however, these are not two free charges: the F1 and D2 content are locked together by the solution and are both fixed by the single charge $n^\psi_{\text{M2}}$.
\section{Fluctuations on the 5-brane}\label{Sec: fluctuations}
\paragraph{Bosons.} In static gauge the scalar degrees of freedom on the brane are associated to the normal directions to the brane. We compute their fluctuations simply by expanding the coordinates
\begin{equation}\label{Eq: fluctuation fields}
	\theta = \bar \theta + \vartheta\,,\qquad x^\mu = \bar x^\mu + w^\mu\,,\qquad B_2 = \bar B_2 + b_2\,,\qquad h_3 = \rmd b_2\,,
\end{equation}
where $\mu = 1,\ldots,4$ labels the directions of $\mathcal M_4$ and $\bar x$ is the equivariant fixed point at which the brane sits. We pair the four scalars along $\mathcal M_4$ into two complex ones,
\begin{equation}\label{Eq: WZ}
	W_1 = w^1 + \rmi w^2\,,\qquad W_2 = w^3 + \rmi w^4\,,
\end{equation}
whose directions correspond to the plane decomposition of the four-dimensional graviphoton flux \eqref{Eq: fixed point flux weights}. We expand the pullback of the metric and the flux in Riemann normal coordinates and fix the Fock-Schwinger gauge for the flux such that
\begin{equation}\label{Eq: FS gauge}
\begin{aligned}
	g_{\mu\nu}(\bar x + Y) =&\, \delta_{\mu\nu} + \mathcal O(Y^2)\,,\\
	A_\mu(\bar x + Y) =&\, \tfrac12 F_{\nu\mu}(\bar x)\, Y^\nu + O(Y^2)\,.	
\end{aligned}
\end{equation}
To expand the bosonic part of the action we note that there is no potential for the scalars in $\mathcal M_4$.\footnote{Indeed, extremising the 5-brane action did not localise it at any particular point in $\mathcal M_4$. The localisation came instead from the $\kappa$-symmetry constraint.} For the $\mathcal \theta$ scalar there is both a mass term and tadpole coming from the DBI and PST actions, which exactly cancel against contributions from the Wess-Zumino terms. Due to the non-trivial background flux the effective worldvolume metric gets contributions from both the DBI and PST terms, and subsequently is not simply the pulled back metric. Instead, the worldvolume fields propagate in an effective background open membrane metric
\begin{equation}
	(\tilde g^{-1}_{6})_{ij} =\left((2K^2-1)-2K^2\sqrt{1-K^{-2}}\right)^{1/6} (C^{-1})_{ij}\,,
\end{equation}
where
\begin{equation}
	(C^{-1})_{ij} = \frac{1}{K}\left(\hat g_{ij} + \frac{1}{4} H_{ikl}H_j^{\phantom{j}kl}\right)\,,\quad K = \sqrt{1 + \frac{1}{24} H_{ijk} H^{ijk}}\,,
\end{equation}
and the worldvolume indices are lowered by the worldvolume metric.\footnote{The tensor $C^{-1}$ is known as the Boillat metric of nonlinear DBI electrodynamics \cite{Gibbons:2000ck}.} This effective metric is independent of $\bar{\theta}$ and has a change in signature from the pulled back metric
\begin{equation}\label{eq:geff-M5}
	\rmd \tilde s^{\,2}_6 = \sum_{i=1}^{3} (\sigma^\psi_i)^2 - \sum_{i=1}^{3}(\sigma^\eta_i)^2\,.
\end{equation}
Finally, the scalars are charged in the normal bundle on the five-brane due to the non-trivial fibration of the graviphoton on the three-spheres \eqref{Eq: fibration of graviphoton over three spheres}. Subsequently, the quadratic action of the scalars takes the form 
\begin{equation}\label{Eq: quadratic bosons}
\begin{aligned}
	S^{(2)}_{\Phi} = \int \rmd^6 \xi \sqrt{|\tilde {g}_6|} \,\left( D_i W_1 D^i \overline{W}_1  + D_i W_2 D^i \overline{W}_2
	 + \partial_i \vartheta \,\partial^i \vartheta \right)\,,
\end{aligned}
\end{equation}
where
\begin{equation}
\begin{aligned}
D \Phi=&\,\rmd \Phi - \rmi q_\Phi\big(  \mathcal A_\psi  + \mathcal A_\eta\big)\Phi\,, \quad \mathcal A_x = \sigma_1^x\,.
\end{aligned}
\end{equation}
The charges and masses of the complex scalars $W_p$ equal
\begin{equation}
\begin{aligned}
q({W_p}) =& \frac12 \left( (1-q_4) f^+ + (-1)^{p+1}(1+q_4)f^- \right)\,.\\
\end{aligned}
\end{equation}
%
The chiral two-form decouples from the scalars. Expanding \eqref{Eq: PST action} to quadratic order in $b_2$ returns the free PST action of a chiral two-form, with the induced metric replaced by the effective one throughout,
\begin{equation}\label{Eq: quadratic two-form}
	S_{b_2}^{(2)} = \int \rmd^6 \xi\, \sqrt{|\tilde g_6|}\left(\frac{1}{2\cdot 3!}\,(\Pi h)_{ijk}(\Pi h)^{ijk} + \frac{\rmi}{4} V_i (\star h)^{ijk} h_{jkl} V^l\right)\,,
\end{equation}
where $\Pi h \equiv h_3 - V\wedge i_V h_3$, and the Hodge star, the PST vector and every index are taken in $\tilde g$. 

The equation of motion of \eqref{Eq: quadratic two-form} is the linearisation of \eqref{Eq: duality constraint}, and reads
\begin{equation}\label{Eq: h selfduality}
	i_V\big(h_3 - \rmi\,\star h_3\big) = 0\,.
\end{equation}
\paragraph{Fermions.}
Upon imposing the $\kappa$-gauge \eqref{eq:kappa-gauge}, and taking the supersymmetric eigenvalue $q_{\text{M5}}=+1$, the Green--Schwarz action equals
\begin{equation}\label{Eq: flux dressed fermion action}
	S_F^{} = \frac{T_{\mathrm{M5}}}{2}\int\rmd^6\xi\sqrt{\det\hat g_6}\, \bar{\Theta}\,m^{ij}\Gamma_j \left(\nabla_i+\frac{1}{24}\Gamma_i\slashed{G}_4 - \frac{1}{8}\slashed{G}_4\Gamma_i\right)\Theta\, 
\end{equation}
and already provides the quadratic contributions to the worldvolume fermions by simply setting the bosonic content to its background value. The kinetic operator provides directly the worldvolume Dirac-operator and additionally a non-trivial connection 
\begin{equation}
	m^{ij} \Gamma_j \nabla_i = \frac{\rmi \rme^{-\rmi \bar\theta}}{L} \left(\slashed{\nabla}_\psi -\rmi \slashed{\nabla}_\eta - \frac{\rmi}{4} \slashed{F} (\rmi \sin^2 \bar \theta \slashed{\mathcal A}_\psi + \cos^2 \bar\theta \slashed{\mathcal A}_\eta)\right)\,.
\end{equation}
The flux-terms provide a similar connection term
\begin{equation}
	m^{ij} \Gamma_j (\frac{1}{24}\Gamma_i\slashed{G}_4 - \frac{1}{8}\slashed{G}_4\Gamma_i) = \frac{\rme^{-\rmi \bar \theta}}{4L}\slashed F \left(\rmi\cos^2\bar\theta\,\slashed{\mathcal A}_\psi + \sin^2\bar\theta\,\slashed{\mathcal A}_\eta\right)
\end{equation}
and adding the two contributions results in a standard Dirac action for the fermions propagating in the effective open membrane metric\footnote{We have canonically normalised the fermions here, which has removed an overall $\rme^{-\rmi \bar \theta}$ dependence.}
\begin{equation}\label{Eq: explicit quadratic fermion action}
\begin{aligned}
	S_F^{(2)} =& \int\rmd^6\xi\sqrt{|\tilde g_6|}\; \bar\Theta \left(\slashed D_\psi^{(\mathfrak q)} -\rmi\slashed D_\eta^{(\mathfrak q)}\right)\Theta\,,
\end{aligned}
\end{equation}
where
\begin{equation}\label{Eq: charged fermion kinetic operators}
	\slashed D_x^{(\mathfrak q)} = \slashed\nabla_x + \mathfrak q \,\slashed{\mathcal A}_x\,,\quad \mathfrak q = \frac14 \slashed{F}\,.
\end{equation}
To get real valued charges we decompose the spinors into the eigenspaces of $\slashed F$
\begin{equation}
	\Gamma_{12}\Theta_{s_{12},s_{34}}=\rmi s_{12}\Theta_{s_{12},s_{34}}\,,
	\qquad
	\Gamma_{34}\Theta_{s_{12},s_{34}}=\rmi s_{34}\Theta_{s_{12},s_{34}}\,,
	\qquad s_{12},s_{34}=\pm1\,,
\end{equation}
such that
\begin{equation}
		\mathfrak q \Theta_{s_{12},s_{34}} = + \rmi Q_{s_{12},s_{34}} \Theta_{s_{12},s_{34}} \,, \qquad Q_{s_{12},s_{34}} = \frac14\left[(s_{12}+s_{34})f^+
	+(s_{12}-s_{34})f^- \right]\,.
\end{equation}
Combining the three fluctuation sectors, the complete quadratic worldvolume action is
\begin{equation}\label{Eq: quadratic action M5 brane}
	S_{\mathrm{M5}}^{(2)}=S_\Phi^{(2)} + S_{b_2}^{(2)} + S_F^{(2)}\,,
\end{equation}
with the three terms given in \eqref{Eq: quadratic bosons}, \eqref{Eq: quadratic two-form}, and \eqref{Eq: explicit quadratic fermion action}. 
\section{The 1-loop partition function}
In this section we will for simplicity restrict our analysis to the empty AdS geometry, with the reason being that the final conclusions will be universal in the backgrounds of minimal gauged $\mathcal N=2$ supergravity. As we will find the one-loop partition function vanishes due to a tower of zero-modes and harmonic fluxes of the chiral two-form which are unaffected by the non-trivial connections for the scalars and fermions. Before continuing to the full 1-loop partition function let us comment on the physical relevance of the final answer. Since we are studying a six-dimensional theory we first need to assert if the 1-loop partition function is logarithmically divergent or not. This divergence is proportional to the trace of the stress tensor which in turn equals the Seeley-De Witt coefficient $b_6$:
\begin{equation}
	\Gamma_{\infty} = -\frac12 \log \frac{L^2}{\Lambda^2} \int \rmd^6 x \sqrt{g} \,b_6\,.
\end{equation}
In the conventions of \cite{Bastianelli:2000hi} this coefficient, for the free $(2,0)$ theory is given by
\begin{equation}
	b_6^{(2,0)} = \frac{1}{(4\pi)^3 7!} (-\frac{245}{8} E_6 - 1680 I_1 - 420 I_2 + 140 I_3 )\,,
\end{equation}
where $E_6$ is the Euler density and the remaining curvature tensors are defined as
\begin{equation}
\begin{aligned}
	I_1 &= C_{k m n l}\, C^{m i j n}\, C_{i}{}^{k l}{}_{j},\\
	I_2 &= C_{k l m n}\, C^{m n i j}\, C_{i j}{}^{k l},\\
	I_3 &= C_{mklp}\left(\nabla^2\delta^m_n + 4R^m{}_n -\frac{6}{5}R\,\delta^m_n\right)C^{nklp}\,,
\end{aligned}
\end{equation}
with the Weyl tensor equal to
\begin{equation}
	C_{ijkl} = R_{ijkl}-\frac{1}{4}\left(g_{ik}R_{jl}-g_{il}R_{jk}-g_{jk}R_{il}+g_{jl}R_{ik}\right) +\frac{R}{20}\left(g_{ik}g_{jl}-g_{il}g_{jk}\right)\,.\\
\end{equation}
To compute these curvature tensors, and the 1-loop partition function of the worldvolume theory below, we introduce as a bookkeeping tool an explicit complex phase in the membrane effective metric on the M5-brane
\begin{equation}\label{Eq: Metric on five brane with radii}
	\rmd \tilde s^2 =  \rmd s_{S^3}^2 - \frac1v \, \rmd s_{\tilde S^3}^2 \,,\quad \text{where} \quad  v = \rme^{-\rmi \varphi},
\end{equation}
and only at the end send $v \to 1$. Note that we have not analytically continued the background AdS$_4 \times S^7/\mathbf{Z}_k$ geometry, we have merely analytically continued the worldvolume theory to keep track of the effects coming from the equal radii but opposite signature spheres.

The Euler density automatically vanishes and the curvature tensors $I_{1,2,3}$ combine to
\begin{equation}\label{Eq: b6 explicit}
	b_6^{(2,0)} = \frac{\big(1 - v\big)^3}{400 \pi^3 }\,,
\end{equation}
which vanishes at $v = 1$. The one-loop determinant is subsequently expected to be logarithmically finite. Note that this would not have been the case if $v = -1$ and we have an all positive signature in the effective metric. Indeed, the logarithmic finiteness is closely related to the similar mechanism for five-branes wrapping an AdS$_3 \times S^3$ inside AdS$_{7} \times S^4$ \cite{Beccaria:2024gkq}. Differently from those setups, we will find here that due to the negative signature in the five-brane effective metric there are infinitely many zero-modes which have to be evaluated carefully and dominate the final M5-brane partition function. To see how these modes contribute we first formally write down the one-loop partition functions of the scalar, Weyl fermion and two form sectors
\begin{equation}
	Z_{\phi}  =  (\text{Det}\,\Delta_0)^{-1/2}\,,\quad Z_{\theta}  =  (\text{Det}\,\Delta_{\frac12})^{1/2}\,.
\end{equation}
and for the 1-loop partition function of the two-form gauge potential we apply a Fadeev-Popov procedure to properly account for the gauge degrees of freedom
\begin{equation}
	\delta_a b_2 = \rmd a_1 \,,\quad \delta_\lambda a_1 = a_1 + \rmd \lambda\,.
\end{equation}
Taking into account the associated ghost fields the one-loop partition function of the two form equals
\begin{equation}\label{Eq: 1-loop for two-form}
	Z_{b_2} = \frac{\text{Det}\,\Delta_{1}}{(\text{Det}\,\Delta_2)^{1/2} (\text{Det}\,\Delta_0)^{3/2}}\,,
\end{equation}
where the fields are all massless, as can be seen in \eqref{Eq: quadratic action M5 brane}, and thus the kinetic operators simply take the form
\begin{equation}
	(\Delta_2)_{kl}^{ij} = -\nabla^2 \delta_{k\ell}^{ij} + 2R_{[k}{}^{[i}\delta_{\ell]}{}^{j]} - R_{kl}{}^{ij}, \quad
(\Delta_1)_{j}^{i} = -\nabla^2 \delta_{j}^{i} + R_{j}^{i}, \quad \Delta_{\frac12} = \rmi \slashed{\nabla}\,,\quad
\Delta_0 = -\nabla^2\,.
\end{equation}
To explicitly evaluate these determinants we follow the same procedure as in Appendix B of \cite{Beccaria:2024gkq} and decompose the determinants into the determinants on the two three-spheres. To keep track of the contributions on the two different spheres we will use the metric as in \eqref{Eq: Metric on five brane with radii} and split the coordinates as $\xi^i \rightarrow (\xi^a ,\xi^{\tilde a})$. Subsequently, the two-form and 1-form kinetic operators organise themselves as
\begin{equation}
\begin{aligned}
	b_{ij} (\Delta_2)_{kl}^{ij} b^{kl} =&\,  b_{ab}(- \nabla^2 + 2)b^{ab} + b_{\tilde a\tilde b}(-\tilde \nabla^2 - 2 v) b^{\tilde a\tilde b}\\
	&+ 2 b_{a\tilde a} \Big(-\nabla^2 -\tilde \nabla^2 + 2\big(1-v\big)\Big) b^{a\tilde a}  \,,\\
	a_i (\Delta_1)^i_j a^j	=&\, a_a(-\nabla^2 + 2 ) a^a + a_{\tilde a} (-\tilde \nabla^2 - 2 v )a^{\tilde{a}}\,,
\end{aligned}
\end{equation}
and so their respective one-loop determinants factorise
\begin{equation}\label{Eq: det factorisation}
\begin{aligned}
	\text{Det}\,\Delta_2 =&\, \text{Det}\,\Delta_{2,0}(2)\,\text{Det}\,\Delta_{1,1}\big(2(1-v)\big)\,\text{Det}\,\Delta_{0,2}(-2v)\,,\\
	\text{Det}\,\Delta_1 =&\, \text{Det}\,\Delta_{1,0}(2)\,\text{Det}\,\Delta_{0,1}(-2v)\,,
\end{aligned}
\end{equation}
where we defined $\Delta_{p,\tilde p}(M) = -\nabla^2 + M$, an operator defined on a field in ${S}^3\times \tilde S^3$ which has $p$-form in ${S}^3$ and $\tilde p$-form in $\tilde S^3$. Before assembling the relevant contributions we note we can simplify the two-form partition function using the fact that in three dimensions we can dualize a two-form to a one-form and subsequently split vector degrees of freedom into transverse and longitudinal components
\begin{equation}
	\text{Det} \,\Delta_{(2,0)} = \text{Det}\,\Delta_{(1,0)} = \text{Det}\,\Delta_{(1\perp,0)}\,\text{Det}\,\Delta_{(0,0)}\,.
\end{equation}
Plugging this into \eqref{Eq: 1-loop for two-form} we find
\begin{equation}\label{Eq: nonchiral two-form}
	Z_{b_2} = \frac{1}{\big(\text{Det}\, \Delta_{1\perp,1\perp}\big(2(1-v)\big)\big)^{1/2}\, \text{Det}\, \Delta_{0,0}(0)}\,.
\end{equation}
We still have to impose that the form is chiral. The determinants in \eqref{Eq: nonchiral two-form} are computed on the fluctuations $h_3 = \rmd b_2$, which are exact. This is in contrary to the harmonic three-form on the worldvolume, which are closed but not exact, and do not enter the Gaussian integral over $b_2$. For a non-chiral two-form this is harmless, because the harmonic modes can be treated as a decoupled flux sum. For a chiral two-form however this is not the case \cite{Witten:1996hc,Belov:2006jd}, and the self-dual partition function is not the square root of the non-chiral one since the flux sum cannot be separated from the Gaussian integral. Instead it factorises as
\begin{equation}\label{Eq: selfdual factorisation}
	Z^{\text{sd}}_{b_2} = \sqrt{Z_{b_2}'}\, \Theta\,,
\end{equation}
where the prime denotes the removal of all harmonic modes,
\begin{equation}\label{Eq: Ng}
	\sqrt{Z_{b_2}'} = \frac{1}{\big(\text{Det}{}'\, \Delta_{1\perp,1\perp}\big(2(1-v)\big)\big)^{1/4}\, (\text{Det}{}'\, \Delta_{0,0}(0))^{1/2}}\,.
\end{equation}
The zero-modes will be discussed in more detail in subsequent sections, where the form of $\Theta$ will also be determined. Combining \eqref{Eq: selfdual factorisation} with the worldvolume scalar and fermion contributions we find that 
\begin{equation}\label{Eq: full 1-loop on worldvolume}
	Z_{\text{1-loop}} = Z_{b_2}^{\text{sd}} Z_{\phi}^5 Z_\theta^{4} = Z_{\text{0-modes}}^{} \Theta  \frac{(\text{Det}{}'\,\slashed D^2)^{1/2}}{\big(\text{Det}{}'\, \Delta_{1\perp,1\perp}\big(2(1-v))\big)\big)^{1/4}\, (\text{Det}{}'\, \Delta_{0,0}(0))^{3}} \,.
\end{equation}
Here $\slashed D$ is the full six-dimensional Dirac operator, with both chiralities included in the determinant.
The Laplace operators of the bosonic fields factorise straight forward
\begin{equation}
\begin{aligned}
	\Delta_{0,0}(0) =&\, \Delta_0 \otimes \mathbf 1 - v \,\mathbf 1 \otimes \tilde \Delta_0 \,,\\
	\Delta_{1\perp,1\perp}\left(2(1-v)\right) =&\,  (-\nabla_1^2) \otimes \mathbf 1 - v \, \mathbf 1 \otimes (-\tilde\nabla_1^2) + 2(1-v) \,,
\end{aligned}
\end{equation}
with eigenvalues and degeneracies also factorising accordingly
\begin{equation}\label{Eq: 6d modes split}
	\Lambda^{(s)}_{\ell\tilde \ell} = \lambda_\ell^{(s)} - v \lambda_{\tilde\ell}^{(s)}\,,\qquad d_{\ell\tilde \ell}^{(s)} = d_\ell^{(s)} d_{\tilde\ell}^{(s)}\,,
\end{equation}
where the three-dimensional eigenvalues and degeneracies are
\begin{equation}
	\lambda_\ell^{(0)} = \ell(\ell+2)\,,\quad d^{(0)}_\ell = (\ell+1)^2\,,\qquad \lambda_\ell^{(1)} = (\ell+2)^2\,,\quad d^{(1)}_\ell = 2(\ell+1)(\ell+3)\,.
\end{equation}
The six-dimensional spinor bundle factorises over the two three-spheres as $S(S^3 \times S^3) \simeq S(S^3) \otimes S(\tilde S^3) \otimes \mathbf{C}^2$, and subsequently the Dirac operator factorises as 
\begin{equation}\label{Eq: Dirac on product}
	\slashed D = \slashed D_1 \otimes \mathbf 1 \otimes \tau_1 + \rmi v^{1/2} \mathbf 1 \otimes \slashed D_2 \otimes \tau_2\,,\qquad
	\slashed D^2 = \slashed D_1^2\otimes\mathbf 1 - v\, \mathbf 1 \otimes\slashed D_2^2\,,
\end{equation}
where $\tau_i$ are anti-commuting Pauli matrices on $\mathbf C^2$. Hence the three-dimensional eigenvalues of the Dirac operator will have to be added in squares
\begin{equation}
	\Lambda^{\frac12}_{\ell\tilde\ell} = \pm \sqrt{\lambda^{(\frac12)}_\ell - v \lambda^{(\frac12)}_{\tilde\ell}}\,, \qquad d_{\ell\tilde\ell}^{(\frac12)} = 4 d^{(\frac12)}_\ell d^{(\frac12)}_{\tilde\ell}\,,
\end{equation}
where
\begin{equation}
	\lambda^{(\frac12)}_\ell = \left(\ell + \frac32\right)^2 \,,\qquad d_\ell^{(\frac12)} = (\ell + 1)( \ell + 2)\,.
\end{equation}
The full 1-loop partition function subsequently simplifies to
\begin{equation}\label{Eq: master sum}
	-\log Z_{\text{1-loop}} = -\log\Theta \;+\; \sum_{s}\sigma_s \sum_{\ell,\tilde\ell\geq0}{} \; d^{(s)}_\ell d^{(s)}_{\tilde\ell}\,
	\log\Big[\lambda^{(s)}_\ell - v\lambda^{(s)}_{\tilde\ell}\Big]\,,
\end{equation}
where $\sigma_s$ keeps track of the degeneracies of bosons and fermions
\begin{equation}
	(\sigma_0,\ \sigma_{\frac12},\ \sigma_{1}) = (3,\ -4,\ \tfrac14)\,.
\end{equation}
For later reference we already record the sum over degeneracies
\begin{equation}\label{Eq: C coefficient}
	D(\ell,\tilde\ell) \equiv \sum_s \sigma_s\, d^{(s)}_\ell d^{(s)}_{\tilde\ell} = -2(\ell+1)(\tilde\ell+1)(\ell+\tilde\ell+2)\,,
	\qquad D(\ell,\ell) = -4(\ell+1)^3\,.
\end{equation}
\subsection{Zero modes}\label{Sec: zero modes}
We treat three zero-mode sectors.  The effective metric \eqref{Eq: Metric on five brane with radii} has split signature at $v=1$, but the modes that then vanish do not all have the same origin.  The harmonic flux sectors and the constant scalar mode require the two special treatments given first.  The third sector is an infinite diagonal tower $\ell=\tilde\ell$ whose eigenvalues cancel between the two spheres; its regulated determinant is evaluated in paragraph (iii).

\paragraph{(i) The flux sectors of the chiral three-form.}
For \(k=1\), the two spheres in \(\mathcal M_6=S^3_\psi\times S^3_\eta\) support quantized
harmonic three-form flux. These sectors are closed but not exact and therefore do not occur in
the Gaussian integral over the two-form potential \(b_2\). A self-dual field instead contributes a separate constrained flux sum \cite{Witten:1996hc,Belov:2006jd}, which we denoted with $\Theta$ in \eqref{Eq: selfdual factorisation}. Fixing the background flux to $n^{\psi}_{\text{M2}}$, the residual sum runs over the harmonic classes
\begin{equation}\label{Eq: flux lattice}
	[n_{\text{M2}}^\psi] = [\bar n_{\text{M2}}^\psi] + r \alpha\,,\qquad r\in \mathbf{Z}\,,
\end{equation}
where $\alpha$ and $\beta$ denote the unit-period harmonic volume forms of $S^3_\psi$ and $S^3_\eta$,
\begin{equation}\label{Eq: alpha beta pairing}
	\int_{S^3_\psi}\alpha = \int_{S^3_\eta}\beta = 1\,,\qquad \int_{\mathcal M_6} \alpha\wedge\beta = 1\,.
\end{equation}
The worldvolume has no torsion or non-trivial three-form gauge potential, and the sum over flux sectors is done without any parity factor. The associated theta characteristic is $[0,0]$ and the flux sectors in \eqref{Eq: flux lattice} are summed with weight $(+1)^r$ rather than $(-1)^r$.

What remains to determine is the Gaussian weight of each flux sector. The Hodge star acts on the harmonic modes as
\begin{equation}\label{Eq: kappa branch}
	\star\alpha = \kappa\,\beta\,,\qquad \star\beta = -\kappa^{-1}\alpha\,,\qquad \star^2 = -1\,, 	\qquad 	\kappa = \rmi\,\rme^{3\rmi\varphi/2}\,,
\end{equation}
where the branch of $\kappa$ is chosen such that it equals $+1$ for fully positive signatures, i.e. at $\varphi = \pi$. With \eqref{Eq: alpha beta pairing} the quadratic form on a general harmonic representative $h = m\alpha + n\beta$ is
\begin{equation}\label{Eq: harmonic action}
	S_{\text{harm}} = \pi\int_{\mathcal M_6} h\wedge\star h
	= \pi\big(\kappa\, m^2 + \kappa^{-1} n^2\big)\,.
\end{equation}
For chiral forms, which are eigenforms of the Hodge-star, this action vanishes. Instead for such fields the weight is the theta function attached to the complex structure that $\star$ induces on $H^3(\mathcal M_6,\mathbf R)$ \cite{Witten:1996hc,Belov:2006jd}. In the gauge $a=\eta_2$ only $\beta$ carries the PST leg, so $i_V\alpha = 0$ while $i_V\beta\neq0$, and the duality constraint \eqref{Eq: h selfduality} becomes 
\begin{equation}
	i_V\big(h_3 - \rmi \star h_3\big)
	= m\, i_V\Big(\big(1+\rmi\,\tau/\kappa\big)\,\alpha + \big(\tau - \rmi \kappa\big)\,\beta\Big)\,,
\end{equation}
where we introduced $\tau=n/m$, which is solved by $(\tau - \rmi\kappa)\,i_V\beta = 0$.\footnote{Had we gauge fixed to $a=\psi_2$ instead it would have been the coefficient of $\alpha$ that survives, giving $1+\rmi\tau/\kappa = 0$ and the same root.} We thus find that
\begin{equation}\label{Eq: period matrix}
	\tau = \rmi\kappa = -\rme^{3\rmi\varphi/2}\,,
\end{equation}
reducing to $\tau = \rmi$ at $\varphi=\pi$. With the characteristic $[0,0]$ the sum over flux sectors is the elliptic $\vartheta$-function
\begin{equation}\label{Eq: Theta}
	\Theta = \sum_{r\in\mathbf{Z}} \rme^{\rmi\pi\tau r^2} = \vartheta_3(0|\tau)\,.
\end{equation}
Since we sum over the lattice associated to the $\alpha$ polarizations this weight is consistent with taking $h = r \alpha$ and $S_{\text{harm}} = \pi\int r\alpha\wedge\star r\alpha = \pi \kappa r^2$, such that the Gaussian weight evaluates to $\rme^{-S_{\text{harm}}} = \rme^{-\pi \kappa r^2} = \rme^{\rmi \pi \tau r^2}$. 

Convergence requires $\operatorname{Im}\tau>0$, i.e.\ $-2\pi/3<\varphi<0$, and the physical configuration is the boundary value of that wedge. In the limit $\varphi\rightarrow0$ the modulus runs into the cusp $\tau\to-1$ and $\Theta$ vanishes exponentially,
\begin{equation}\label{Eq: theta expansion}
	-\log \Theta = \frac{\pi}{6 |\varphi|} + \frac12 \log \frac{3|\varphi|}{2}
	+ \frac{\pi \rmi}{8} - \log 2 + \mathcal O(\varphi)\,.
\end{equation}
This contribution is independent of the four-dimensional deformations giving a connection to the scalars and fermions, which is the reason that this M5-brane partition function vanishes irrespective of those deformations.
\paragraph{(ii) The honest scalar zero modes at $\ell=\tilde\ell=0$.}
These are the constant modes, which have to be treated with collective coordinates. These modes only exist in the scalar sector
\begin{equation}
	\lambda^{(0)}_0 = 0\,,\qquad \lambda^{(\frac12)}_0 = \tfrac94\,,\qquad \lambda^{(1)}_0 = 4\,.
\end{equation}
Since these modes are constant on both the three-spheres the relative signature does not influence the following analyisis making it valid for any value of the phase $v$. The degeneracy of the scalar mode is $d^{(0)}_0 d^{(0)}_0 = 1$ and its weight in \eqref{Eq: master sum} is $\sigma_0 = 3$, made up of $\tfrac52$ from the five worldvolume scalars and $\tfrac12$ from the longitudinal two-form determinant. 

Four of the scalars correspond to the position $\bar x^\mu$ of the brane in $\mathcal M_4$. These modes do not need to be integrated in collective coordinates, instead their contributions are determined directly through the equivariant localization of the brane in the $\mathcal M_4$, as was explained in \cite{Gautason:2025per} for M2-brane instantons. In the case of empty AdS with a spherical boundary there is one fixed point and these modes simply contribute $1$ to the partition function.\footnote{For other backgrounds the equivariant fixed points were given in Section \ref{Sec: backgrounds}.} The fifth scalar zero-mode is the polar angle $\bar\theta$. It does not have a potential and one could think that it is to be integrated in collective coordinates. We fixed however the charge on the worldvolume of the brane $n_{\text{M2}}^{\psi}$, which subsequently also fixes the position $\bar \theta$ of the brane. 

%
\paragraph{(iii) The cancelling modes between the two spheres.}
This family is an artefact of the split signature.  At $v=1$, every mode with $\ell=\tilde\ell$ has zero six-dimensional eigenvalue, even though it is an ordinary massive mode of each three-sphere operator. These modes cannot be treated with collective coordinates. Instead, we compute their contribution as a function of the phase $v$ and put $v=1$ at the end of the calculations. The eigenvalues of the cancelling modes are then
\begin{equation}\label{Eq: diagonal eigenvalue}
	\Lambda^{(s)}_{\ell\ell}=(1-v)\lambda^{(s)}_\ell\,,
\end{equation}
and define the following $\zeta$-functions
\begin{equation}\label{Eq: diagonal spectral zeta}
	\zeta_s(w)=\sum_\ell{}\big(d^{(s)}_\ell\big)^2
	\big(\lambda^{(s)}_\ell\big)^{-w}\,,
\end{equation}
where the scalar $\ell=0$ mode treated in paragraph (ii) is excluded.  The
three formal sums we have to compute are subsequently
\begin{equation}\label{Eq: explicit diagonal spectral zetas}
\begin{aligned}
	\zeta_0(w) &= \sum_{\ell=1}^{\infty} \frac{(\ell+1)^4}{\big(\ell(\ell+2)\big)^w}\,,\\
	\zeta_{1/2}(w) &= \sum_{\ell=0}^{\infty} \frac{(\ell+1)^2(\ell+2)^2}{(\ell+\frac32)^{2w}}\,,\\
	\zeta_1(w) &= 4\sum_{\ell=0}^{\infty} \frac{(\ell+1)^2(\ell+3)^2}{(\ell+2)^{2w}}\,,
\end{aligned}
\end{equation}
Their values and first derivatives at $w=0$ are
\begin{equation}\label{Eq: sector counts}
\begin{aligned}
	\zeta_0(0)&=-1\,,
	&\qquad \zeta_0'(0)
	&=2\zeta'(-4)+12\zeta'(-2)-\log\pi\,,\\
	\zeta_{1/2}(0)&=0\,,
	&\zeta_{1/2}'(0)
	&=-\frac{15}{8}\zeta'(-4)+\frac34\zeta'(-2)
	-\frac1{16}\log2\,,\\
	\zeta_1(0)&=-2\,,
	&\zeta_1'(0)
	&=8\zeta'(-4)-16\zeta'(-2)-4\log(2\pi)\,.
\end{aligned}
\end{equation}
With the weights in \eqref{Eq: master sum}, we find the diagonal contribution to the one-loop free energy 
\begin{equation}\label{Eq: cancelling modes total}
\begin{aligned}
	-\log Z_{\mathrm{diag}} &= \sum_s\sigma_s\sum_\ell{}'\big(d^{(s)}_\ell\big)^2
	\log\!\big[(1-v)\lambda^{(s)}_\ell\big]\\
	&= -\sum_s\sigma_s\zeta_s'(0) +\sum_s\sigma_s\zeta_s(0)\log(1-v)\\
	&= C_{\text{diag}}-\frac72\log(1-v)\,, 
	\qquad
	C_{\text{diag}} =\frac34\log2+4\log\pi-\frac{31}{2}\zeta'(-4)-29\zeta'(-2)\,.
\end{aligned}
\end{equation}
\subsection{The non-zero modes}\label{Sec: nonzero modes}
What remains is to compute the modes with $\ell\neq\tilde\ell$, for which we can safely set $v=1$ and factor the eigenvalue in \eqref{Eq: master sum} as
\begin{equation}
	\log\Lambda^{(s)}_{\ell\tilde\ell}\Big|_{\ell\neq\tilde\ell} = \log(\ell-\tilde\ell) + \log(\ell+\tilde\ell+c_s)\,,
\end{equation}
where
\begin{equation}
	(c_0,c_{\frac12},c_1) = (2,\ 3,\ 4)\,.
\end{equation}
We will first evaluate the  $\log(\ell+\tilde\ell+c_s)$ sums and let $\ell,\tilde\ell$ also run over the diagonal values, and only in the end compensate for this overcounting. This gives three sums, one per sector,
\begin{equation}\label{Eq: N sums}
\begin{aligned}
	\mathcal S_{(0)} &= \sum_{\ell,\tilde\ell\geq0}(\ell+1)^2(\tilde\ell+1)^2 \log\!\big[\ell+\tilde\ell+2\big]\,,\\
	\mathcal S_{(1/2)} &= \sum_{\ell,\tilde\ell\geq0}(\ell+1)(\ell+2)(\tilde\ell+1)(\tilde\ell+2) \log\!\big[\ell+\tilde\ell+3\big]\,,\\
	\mathcal S_{(1)} &= 4\sum_{\ell,\tilde\ell\geq0}(\ell+1)(\ell+3)(\tilde\ell+1)(\tilde\ell+3) \log\!\big[\ell+\tilde\ell+4\big]\,.
\end{aligned}
\end{equation}
Each sum is evaluated by reorganising it in terms of
$n=\ell+\tilde\ell$, and including a polynomial weight
$Q(n)=\sum_jq_j(n+c_s)^j$
\begin{equation}\label{Eq: 2d zeta}
	\sum_{\ell,\tilde\ell\geq0} d_{\ell,\tilde \ell}\log\big[\ell+\tilde\ell+c_s\big] = -\sum_j q_j\, \zeta'(-j,c)\,,
	\qquad \zeta'(-j,c) = \zeta'(-j) + \sum_{i=1}^{c-1} i^{\,j}\log i\,,
\end{equation}
so that everything reduces to $\zeta'(-j)$ and logarithms of integers. Explicitly
\begin{equation}\label{Eq: N closed}
\begin{aligned}
	\mathcal S_{(0)} &= \frac{1}{30}\Big(\zeta'(-1)-\zeta'(-5)\Big)\,,\\
	\mathcal S_{(1/2)} &= -\frac{1}{30}( 4 \zeta'(-1) - 5 \zeta'(-3) + \zeta'(-5))\,,\\
	\mathcal S_{(1)} &= -2\log 2\pi - \frac{1}{30} (76 \zeta'(-1) + 120 \zeta'(-2) - 80 \zeta'(-3) + 4\zeta'(-5))\,,
\end{aligned}
\end{equation}
and summing them with appropriate degeneracies we find that all odd valued $\zeta$-functions vanish
\begin{equation}\label{Eq: N total}
	\mathcal S_{\mathrm{all}} = \sum_s \sigma_s \mathcal S_{(s)}
	= -\frac12\log 2\pi - \zeta'(-2)\,.
\end{equation}
We still have to correct for the diagonal terms, which contribute
\begin{equation}\label{Eq: N diagonal}
	\mathcal S_{\mathrm{diag}}
	=\sum_s\sigma_s\sum_\ell
	\big(d_\ell^{(s)}\big)^2\log(2\ell+c_s)
	=-\frac18\log2+\frac12\log\pi-\frac{31}{4}\zeta'(-4)
	+\frac72\zeta'(-2)\,.
\end{equation}
Subtracting \eqref{Eq: N diagonal} removes every diagonal term, including the spurious scalar $(0,0)$ term introduced by the extension.  The genuine off-diagonal $\log(\ell+\tilde\ell+c_s)$ contribution is therefore
\begin{equation}\label{Eq: N offdiagonal}
	\mathcal S = \mathcal S_{\mathrm{all}} - \mathcal S_{\mathrm{diag}} =-\frac38\log2-\log\pi+\frac{31}{4}\zeta'(-4) - \frac92\zeta'(-2)\,.
\end{equation}
Finally, the off-diagonal $\log(\ell-\tilde\ell)$ contribution is
\begin{equation}\label{Eq: f0 f1}
	\sum_{\ell\neq\tilde\ell} D(\ell,\tilde\ell)\log(\ell-\tilde\ell)
	=2f_1-\rmi\pi f_0\,,
	\quad
	f_1=-\frac{1}{60}\big(\log2\pi+10\zeta'(-2)\big)\,,
	\quad
	f_0=\frac{1}{60}\,,
\end{equation}
Combining its real part with \eqref{Eq: N offdiagonal} gives the finite off-diagonal determinant
\begin{equation}\label{Eq: offdiagonal finite}
	C_{\text{finite}} = \mathcal S + 2f_1 - \rmi \pi f_0 = - \frac{49}{120}\log2 - \frac{31}{30}\log\pi + \frac{31}{4}\zeta'(-4) - \frac{29}{6}\zeta'(-2) - \frac{\rmi \pi}{60}\,.
\end{equation}
\subsection{The full one-loop partition function}\label{Sec: consolidation}
We can now combine the flux sum, the collective-coordinate measure, and the diagonal and off-diagonal determinants. The complete regulated one-loop partition function is 
\begin{equation}\label{Eq: consolidated}
\begin{aligned}
	Z_{\text{1-loop}} &= Z_{\text{0-modes}}\,\Theta \big(1-\rme^{-\rmi\varphi}\big)^{7/2} \exp\!\left[-C_{\text{diag}} - C_{\text{finite}}\right] = \mathcal N  |\varphi|^3 \rme^{-\frac{\pi}{6 |\varphi|}} \big[1+\mathcal O(\varphi)\big]\,,
\end{aligned}
\end{equation}
where
\begin{equation}
	\mathcal N = \rme^{\frac{17 \pi \rmi}{120}} \sqrt{\frac{2}{3}} \exp[\frac{79}{120} \log 2 - \frac{89}{30} \log \pi + \frac{31}{4} \zeta'(-4) + \frac{203}{6} \zeta'(-2)]\,.
\end{equation}
The important result is that in the limit where $\varphi \rightarrow 0$ we find that the 1-loop partition function vanishes. The suppression is exponential in $\varphi$, due to the sum over harmonic three-form fluxes, and thus dominates any effect coming from the infinite number of cancelling modes due to the opposite sign signatures on the two three-spheres. Only two factors in \eqref{Eq: consolidated} depend on $\varphi$: the theta function, which behaves as $|\varphi|^{-1/2}\rme^{-\pi/6|\varphi|}$ by \eqref{Eq: theta expansion}, and $(1-\rme^{-\rmi\varphi})^{7/2}\sim|\varphi|^{7/2}$ from the diagonal tower, the exponent $7/2$ being $-\sum_s\sigma_s\zeta_s(0)$ from \eqref{Eq: cancelling modes total}. Their product is the $|\varphi|^{3}\rme^{-\pi/6|\varphi|}$ in \eqref{Eq: consolidated}. It is worth noting that the diagonal modes are not zero modes of any symmetry, they are an artefact of the split signature so unlike the sectors in (i) and (ii) there is are collective coordinates to deal with them.
\subsection{Generalisation to \texorpdfstring{$k>1$}{k > 1}}\label{Sec: k larger than one}
So far we have kept the orbifold number $k=1$. The orbifold acts freely on the worldvolume coordinates of the brane, explicitly:
\begin{equation}\label{Eq: orbifold on worldvolume}
	\quad \psi_1 \rightarrow \psi_1 + \frac{2\pi}{k}\,,\qquad \eta_1 \rightarrow \eta_1 - \frac{2\pi}{k}\,,
\end{equation}
The worldvolume $\mathcal M_6 = (S^3_\psi\times S^3_\eta)/\mathbf Z_k$ is thus smooth and the quotient map is a $k$-fold cover. Local quantities such as the effective metric and the kinetic operators of the worldvolume fields are untouched. There are some differences though: the spectrum of modes are projected down to the $\mathbf Z_k$-invariant ones, the volume of $\mathcal M_6$ is reduced by $k$, and the lattice of chiral harmonic modes over which one is to sum changes.

The projection onto the $\mathbf Z_k$-singlet modes can be done by replacing the degeneracy $d^{(s)}_\ell d^{(s)}_{\tilde\ell}$ for each field by
\begin{equation}\label{Eq: projector}
	P_s(\ell,\tilde\ell) = \frac1k\sum_{j=0}^{k-1}\chi_s(\ell;\kappa_j)\,\chi_s(\tilde\ell;-\kappa_j)\,, \qquad \kappa_j = \frac{2\pi j}{k}\,,
\end{equation}
where $\chi_s(\ell;\kappa)$ is the character of the $\mathbf{Z}_k$-action on the level-$\ell$ eigenspace of the corresponding kinetic operator. Because \eqref{Eq: orbifold on worldvolume} rotates the two spheres oppositely. Writing $\sigma^x_2 \mp \rmi \sigma^x_3 \rightarrow \rme^{\pm 2\rmi a}(\sigma^x_2 \mp \rmi\sigma^x_3)$ for a shift of $x_1$ by $a$, the generator acts on the level-$\ell$ eigenspaces as an $SU(2)$ element of angle $\theta_j$ together with the induced rotation of the left-invariant frame, and
\begin{equation}\label{Eq: characters}
\begin{aligned}
	\chi_0(\ell;\kappa) =&\, (\ell+1)\frac{\sin(\ell+1)\kappa}{\sin\kappa}\,,\\
	\chi_{1/2}(\ell;\kappa) =&\, \frac12\left[(\ell+2)\frac{\sin(\ell+1)\kappa}{\sin\kappa}+(\ell+1)\frac{\sin(\ell+2)\kappa}{\sin\kappa}\right]\,,\\
	\chi_{1}(\ell;\kappa) =&\, (\ell+3)\frac{\sin(\ell+1)\kappa}{\sin\kappa}+(\ell+1)\frac{\sin(\ell+3)\kappa}{\sin\kappa}\,,
\end{aligned}
\end{equation}
each reducing to $d^{(s)}_\ell$ as $\kappa\rightarrow 0$, so that $P_s \rightarrow d^{(s)}_\ell d^{(s)}_{\tilde\ell}$ at $k=1$. In the vector and the fermion sectors both towers of the three-dimensional operator have to be retained: the two helicities of the transverse vector, and the two signs of the three-dimensional Dirac operator, which enter the six-dimensional eigenvalue only through $\lambda^{(s)}_\ell$ and which carry the two characters $(\ell+1)\sin(\ell+2)\kappa/\sin\kappa$ and $(\ell+2)\sin(\ell+1)\kappa/\sin\kappa$ separately. The five worldvolume scalars are singlets under $\mathbf{Z}_k$ and projected with $\chi_0$. Since the quotient is free there are no twisted sectors to add.

With \eqref{Eq: projector} the one-loop partition function \eqref{Eq: master sum} becomes
\begin{equation}\label{Eq: master sum k}
	-\log Z_{\text{1-loop}} = -\log\Theta_k \;+\; \sum_{s}\sigma_s \sum_{\ell,\tilde\ell\geq0} \; P_s(\ell,\tilde\ell)\,
	\log\Big[\lambda^{(s)}_\ell - v\lambda^{(s)}_{\tilde\ell}\Big]\,,
\end{equation}
with the same weights $\sigma_s$ and the same eigenvalues as before. The projection changes multiplicities only: it does not move eigenvalues, so the diagonal tower $\ell=\tilde\ell$ of paragraph (iii) is still exactly the set of modes that degenerate at $v=1$, and there are still no fermionic zero modes. It contributes a term 
\begin{equation}
	Z_{\text{diag},k}\sim (1-e^{-i\varphi})^{\frac72}\,.
	\label{eq:diag-k}
\end{equation}

The scalar zero-modes and finite part of the one-loop partition function can be treated in the same way as in the case of $k=1$. The question is whether the sum over harmonic sectors of the chiral two-form still makes the partition function vanish or not. To answer this we note that in addition to the two three-spheres the quotient contains the $\mathbf{Z}_k$-invariant cycle $\mathcal C_{\text{NS}}$ of \eqref{Eq: CNS}. 
The harmonic representatives $\alpha$ and $\beta$ of section \ref{Sec: zero modes} integrate to
\begin{equation}\label{Eq: lens periods}
	\int_{\mathcal C_{\text{NS}}}\alpha = \frac1k\,,\qquad \int_{\mathcal C_{\text{NS}}}\beta = \frac1k\,,
\end{equation}
neither being an integral class for $k>1$. The integral classes are then instead
\begin{equation}\label{Eq: k lattice}
	H^3(\mathcal M_6,\mathbf Z) \;\simeq\; \big\{\,a\,\alpha + b\,\beta\ :\ a,b\in\mathbf Z\,,\ a+b\equiv 0\ \text{mod}\ k \,\big\}\,.
\end{equation}
We introduce the integral symplectic basis, whose integer combinations automatically satisfy the constraint in \eqref{Eq: k lattice}:
\begin{equation}
	e_1\equiv \alpha-\beta\,,\qquad e_2\equiv k\beta\,,\qquad \int_{\mathcal{M}_6}e_1\wedge e_2 =1\,.
\end{equation}
The chiral flux sum runs over $\mathbf Z e_1$, with $\mathbf Z e_2$ providing the complementary lattice. A general self-dual harmonic three-form in such a basis reads 
\begin{equation}
	h= m (\alpha + \tau \beta) = m(e_1 + \tau_k e_2)\,,\qquad \tau_k = \frac{1+\tau}{k}\,.
\end{equation} 
In the new basis, we take the associated theta characteristic being $[0,\frac12]$.\footnote{Naively one could instead take the characteristic $[0,0]$ in the basis $(e_1,e_2)$, but this is inconsistent: it would give $\Theta_k=\vartheta_3(0|\tau_k)$, which at $k=1$ is $\vartheta_3(0|1+\tau)=\vartheta_4(0|\tau)$ and not the $\vartheta_3(0|\tau)$ found in \eqref{Eq: Theta}. That choice would in any case give $\Theta_k\simeq\sqrt{2k/3|\varphi|}$ and hence $Z_{\text{1-loop}}\propto|\varphi|^3$, so the vanishing is unaffected.}
The sum over flux sectors is now the elliptic $\vartheta$-function \cite{Belov:2006jd}
\begin{equation}
	\Theta_k  = \vartheta
	\begin{bmatrix}
	0\\
	\frac12
	\end{bmatrix}
	(0|\tau_k) = \sum_{r\in\mathbf{Z}}e^{\,i\pi\tau_kr^2+i\pi r}=\vartheta_4(0|\tau_k).
\end{equation}
It reduces to \eqref{Eq: Theta} when $k=1$ since $\vartheta_4(0|1+\tau)=\vartheta_3(0|\tau)$. Recall that $\tau=i\kappa=-e^{3i\varphi/2}$, and thus in the limit $\varphi\rightarrow 0^-$ we get 
\begin{equation}\label{Eq: theta-k expansion}
	-\log \Theta_k = \frac{k\pi }{6 |\varphi|} + \frac12 \log \frac{3|\varphi|}{2k}
	+ \frac{k\pi \rmi}{8} - \log 2 + \mathcal O(\varphi)\,.
\end{equation}
and hence 
\begin{equation}
	|\Theta_k| \sim 2\sqrt{\frac{2k}{3|\varphi|}}e^{-\frac{k\pi}{6|\varphi|}}\,.
\end{equation}
Together with the zero-modes contribution from the diagnoal terms \eqref{eq:diag-k}, it turns out that the 5-brane instanton partition function vanishes at $\varphi \rightarrow 0$ for any $k$.

The conclusion of Section \ref{Sec: consolidation} is thus unchanged for all $k$. We find evidence that the full M-theory partition function in the asymptotically locally AdS backgrounds of minimal $\mathcal N=2$ gauged supergravity is to be expanded into saddles of M2-branes. The non-trivial supersymmetric M5-brane embeddings are those with non-trivial worldvolume flux, and their origin comes from polarized not M2-branes. This is in contrast to the M5-brane giant graviton expansion of the ABJM superconformal index, which is an expansion at fixed $N$ \cite{Arai:2020uwd,Gaiotto:2021xce,Beccaria:2023cuo}. Our results instead suggest that the bulk partition function, expanded at fixed $\mu$ takes the form
\begin{equation}
	\mathcal Z_{\text{M}}(\mu) = {-S_{\text{SUGRA}}(\mu)} + \ldots + \sum_{\text{M2s}} \rme^{-S_{\text{M2}}(\mu)} (Z_{\text{1-loop}}^{\text{M2}} + \ldots) \,,\quad \mu \sim (L/\ell_p)^3
\end{equation}
where the only non-trivial M5-brane instantons arise from stacks of M2-brane instantons for which $S_{\text{M2}} \sim n \mu $ with the instanton level itself scaling as $n \sim \mu$. Upon quantising our set of M5-instantons we further find that their one-loop partition function vanishes, trivialising their contribution.

\section*{Acknowledgements}
We are thankful to Fridrik Gautason, Joel Karlsson, and Arkady Tseytlin for useful discussions. We also thank Arkady Tseytlin for useful comments on the draft. JvM is supported by the STFC Consolidated Grant ST/X000575/1. JvM is grateful for the continuous hospitality of the ITF at the KU Leuven. ZW is grateful to J. Minahan   for hospitality at Uppsala University while this work was in its final stages.

\appendix
\section{\texorpdfstring{Analytic continuation from AdS$_7 \times S^4$}{Analytic continuation from AdS7 x S4}}
In this appendix we show that the Euclidean M5-brane instanton discussed in section 3 is related, through a continuation of the complexified eleven-dimensional fields, to the Lorentzian $AdS_3\times S^3$ defect M5-brane in $AdS_7\times S^4$.
 The  $AdS_{7}\times S^{4}$ metric and flux in in Lorentzian signature are 
\begin{align}
&\rmd s_{  AdS_{7}\times S^{4}}^2=L^2(\rmd\theta^2+\cosh^2\theta\,ds^2_{AdS_3}+\sinh^2\theta\,\rmd s^2_{S^3})+\frac{L^2}{4}\rmd s^2_{S^4}\ , \qquad F_4=\frac{3}{8}L^3\text{vol}_{S^4}
\end{align}
where we explicitly each part of the metric
\begin{align}
&\rmd s^2_{AdS_3}=\rmd\eta_{1}^{2}+\rmd\eta_2^{2}+\rmd\eta_{3}^{2}-2\cosh(2\eta_2)\rmd\eta_1\rmd\eta_3\\
&\rmd s^2_{S^4}=\rmd u^2+\sin^2 u\left(\cos ^2 \rho \rmd \phi_1^2+\rmd \rho^2+\sin ^2 \rho \rmd \phi_2^2\right)\\
&\text{vol}_{S^4}=\sin^3u\cos\rho\sin\rho \rmd u\wedge \rmd\rho\wedge \rmd\phi_1\wedge \rmd\phi_2\label{vols4}
\end{align}
Define the analytic continuation
\begin{equation}
L\rightarrow 2L,\quad \theta\rightarrow i\theta,\quad \eta_1 \rightarrow i\eta_1,\quad \eta_2 \rightarrow \frac\pi4+ i\eta_2,\quad \eta_ 3\rightarrow i\eta_3,\quad u\rightarrow iu.
\end{equation}
The resulting metric and flux become
\begin{align}
&-\rmd s_{  H^{4}\times S^{7}}^2=-4L^2(\rmd\theta^2+\cos^2\theta\,\rmd s^2_{S^3}+\sin^2\theta\,\rmd s^2_{S^3})-{L^2}\rmd s^2_{AdS_4}\ ,\qquad F_4={3}L^3\text{vol}_{AdS_4}\ .
\end{align}
We obtain a minus sign metric 11d supergravity solution through analytic continuation.

\subsection{$AdS_3 \times S^3\in AdS_7 \in AdS_7 \times S^4$ solution}  
Take the static gauge $\xi^i=(\psi_1,..,\psi_3,\eta_1,...,\eta_3),\;i=0,...,5$, and gauge choice $a=\eta_2$.  The ansatz for the world-volume three form field strength is
\begin{equation}
H_3=bL^3\text{vol}_{S^3}
\end{equation}
Minimize the action with respect to $\theta$ to find the position of the brane.  The induced metric and the 2-from are 
\begin{equation}
\sqrt{-|G|}=L^6\cosh^3\theta \sinh^3 \theta\sqrt{-g_A}\sqrt{g_S}
\end{equation}
\begin{equation}
\widetilde {\mathrm{H}}^{i j}=\frac{1}{6 \sqrt{-|G|}} \frac{1}{\sqrt{-(\partial a)^2}} \varepsilon^{i j k \ell m n} \partial_k a \mathrm{H}_{\ell m n},\quad \widetilde {\mathrm{H}}_{02}=ibL^2\frac{\cosh^2 \theta}{\sinh^3 \theta }\sqrt{-g_A}
\end{equation}
The volume part of the Lagrangian is then 
\begin{equation}
L_1=-\sqrt{-|G_{ij}+\hat{\mathrm{H}}_{ij}|}
=-L^6\sqrt{-g_A}\sqrt{g_S}\cosh^3\theta\sqrt{b^2+\sinh^6\theta}
\end{equation}
There is also the $C_6$ part of the Lagrangian
\begin{equation}
L_2=C_6=\frac{1}{32}L^6\sqrt{-g_A}\sqrt{g_S}(-9\cosh2\theta+\cosh 6\theta+8)
\end{equation}
Rewriting the variables as $x=\cosh \theta$ and $y=\cosh2\theta$,
\begin{equation}
L_1+L_2=-\frac{1}{8}\sqrt{8(1+y)^3b^2+(y^2-1)^3}+\frac{1}{8}(y^3-3y+2)
\end{equation}
This is extremized by
\begin{equation}
y=1+2{b}=\cosh2\theta,\quad b=\sinh^2\theta=\kappa^2.
\end{equation}
The solution is summarized as following
\begin{equation}
\theta=\theta_0,\quad \sinh \theta_0=\kappa,\quad \rho=0,\quad H_3=\kappa^2L^3\text{vol}_{S^3}\ .
\end{equation}

\label{Appendix: Analytic continuation}

\section{The Killing spinor for $\mathrm{AdS}_4\times S^7/\mathbb Z_k$ background}
\label{appD}
In this section we write down the Killing spinor that solve the Killing spinor equation \be
\delta \Psi_M={D}_M \epsilon \equiv\left(\partial_M-\Omega_M\right) \epsilon=0,
\qquad 
\delta {D}_M= \left( \nabla_M + \frac{1}{24} \Gamma_M \slashed{G}_4 - \frac18 \slashed{G}_4 \Gamma_M \right) 
\ee
for the $\mathrm{AdS}_4\times S^7$ background. The Killing spinor equation separates into two independent parts describing the 4d and 7d equations respectively. We decompose the gamma matrices as $\Gamma_\mu =i\gamma_{(4)}\gamma_\mu\otimes \mathbb 1, \;\Gamma_i=-\gamma_{(4)}\otimes \Sigma_i$ and the Killing spinor as $\epsilon=\eta\otimes \chi$.

The 7d Killing spinor is 
\begin{align}
&\label{D12}
\chi=e^{ \Omega_{\theta }\theta}
e^{\Omega_{\psi_1}\psi_1}
e^{\Omega_{\psi_2}\psi_2}
e^{\Omega_{\psi_3}\psi_3}
e^{\Omega_{\eta_1}\eta_1}
e^{\Omega_{\eta_2}\eta_2}
e^{\Omega_{\eta_3}\eta_3}
\chi_0.\\
&\label{D13}
\Omega_\theta=\tfrac i2  \Sigma_{{5}},\quad
 \Omega_{\psi_1}=\tfrac{1}2(\Sigma_{56}+\Sigma_{78}),\quad  \Omega_{\psi_2}=\tfrac{1}2(\Sigma_{57}-\Sigma_{68}),\quad
 \Omega_{\psi_3}=\tfrac{1}2(\Sigma_{67}+\Sigma_{58})\\
&\label{D14}
 \Omega_{\eta_1}=\tfrac{1}2(\Sigma_{10\,11}+i\Sigma_{9}),\quad  \Omega_{\eta_2}=\tfrac{1}2(-\Sigma_{9\,11}+i\Sigma_{10}),\quad
 \Omega_{\eta_3}=\tfrac{1}2(\Sigma_{9\,10}+i\Sigma_{11}).
\end{align}
For the AdS$_4 \times S^7/\mathbb Z_k$ background, the local Killing spinor equation is unchanged by the quotient.
There is additional global condition on the  invariance of the Killing spinor under the identification
$
(\psi_1,\eta_1) \sim \big( \psi_1 + \frac{2\pi}{k}, \eta_1 - \frac{2\pi}{k} \big)
$
:
\al{
\exp\left[\frac{2\pi}{k}\left(
\Omega_{\psi_1}-\Omega_{\eta_1}\right)\right]\chi_0=\exp\left[\frac{\pi}{k}\left(\Sigma_{56}+\Sigma_{78}-\Sigma_{10\,11}-i\Sigma_9\right)\right]\chi_0
=\chi_0 .\label{E5}
}
Here $\chi_0$ is an 8-component seven-dimensional constant spinor. Decomposing the constant spinor by projection conditions 
\al{
&e^{\frac{\pi}{k}\Sigma_{56}}\chi_{0,s_1,s_2,s_3}=s_1 \chi_{0,s_1,s_2,s_3}\ , \quad 
e^{\frac{\pi}{k}\Sigma_{78}}\chi_{0,s_1,s_2,s_3}=s_2 \chi_{0,s_1,s_2,s_3}\ , \nonumber\\
&
e^{-\frac{\pi}{k}\Sigma_{10\,11}}\chi_{0,s_1,s_2,s_3}=s_3 \chi_{0,s_1,s_2,s_3}\ , \quad 
s_1,s_2,s_3=\pm\ ,
}
the conditions on the spinor components are 
\al{
e^{\tfrac{ 4\pi}k}\chi_{0,+++}=\chi_{0,+++}\ ,\quad e^{-\tfrac{ 4\pi}k}\chi_{0,---}=\chi_{0,---}\ ,\label{E6}
}
and the permutations of 
\al{&
\chi_{0,-++}=\chi_{0,-++}\ ,\quad \chi_{0,+-+}=\chi_{0,-++}\ ,\quad \chi_{0,++-}=\chi_{0,++-}\ ,\quad\nonumber\\
& \chi_{0,--+}=\chi_{0,--+}\ ,\quad \chi_{0,-+-}=\chi_{0,-+-}\ ,\quad \chi_{0,+--}=\chi_{0,+--}\ .
}
For $k=1,2$, (\ref{E6}) is trivially satisfied; all the spinor components are preserved. For $k>2$, the condition in (\ref{E6}) is satisfied with the components $\chi_{0,+++}=\chi_{0,---}=0$, projecting out $1/4$ of the components. 
In 11d, 32 supercharges are preserved for $k=1,2$ and 24 supercharges are preserved for $k>2$.

\bibliography{M5BIB.bib}
\bibliographystyle{JHEP}

\end{document}